\documentclass[11pt]{article}
\usepackage{graphicx}
\usepackage{amsmath}
\usepackage{cite}

\newcommand{\BABARPubYear}    {08}

\newcommand{\BABARConfNumber} {014}
\newcommand{\SLACPubNumber}   {13326}
\newcommand{\LANLNumber}      {0807.4226}

\input babarsym.tex

\def\sPlots{\ensuremath{\hbox{$_s$}{\cal P}lots}\xspace}
\def\Abar  {\kern 0.2em\overline{\kern -0.2em A}{}\xspace}

\def\Btopipi   {\ensuremath{\B \to \pi\pi}\xspace}
\def\Btopipiz   {\ensuremath{\Bpm \to \pipm\piz}\xspace}

\def\Bztopizpiz   {\ensuremath{\Bz \to \piz\piz}\xspace}

\def\Bztopipi   {\ensuremath{\Bz \to \pipi}\xspace}
\def\Bztopippim   {\ensuremath{\Bz \to \pip\pim}\xspace}
\def\Bztokpi   {\ensuremath{\Bz \to \Kp\pim}\xspace}

\def\Bztohh      {\ensuremath{\Bz\to h^+h^{\prime -}}\xspace}

\def\Bztokzpiz   {\ensuremath{\Bz \to \Kz\piz}\xspace}

\def\Bptorhoppiz   {\ensuremath{\B^+ \to \rho^+\piz}\xspace}

\def\Btag {\ensuremath{B_{\rm tag}}}
\def\Brec {\ensuremath{B_{\rm rec}}}
\def\Bflav {\ensuremath{B_{\rm flav}}}

\def\alphaeff {\ensuremath{\alpha_{\rm eff}}\xspace}
\def\de {\ensuremath{\Delta E}\xspace}
\def\cossph   {\ensuremath{|\cos{\theta_{\scriptscriptstyle S}}|\;}\xspace}
\def\fish    {\ensuremath{\cal F}\xspace}

\def\thetac {\ensuremath{\theta_{\rm C}}\xspace}

\def\sss{\scriptscriptstyle}
\def\barpd{{\raise.35ex\hbox
{${\sss (}$}}--{\raise.35ex\hbox{${\sss )}$}}}
\def\BorBbar{\hbox{$B^{0}$\kern-1.25em\raise1.5ex\hbox{\barpd}}}

\def\qq {\ensuremath{q\bar{q}}\xspace}
\def\spipi {\ensuremath{S_{\pi\pi}}\xspace}
\def\cpipi {\ensuremath{C_{\pi\pi}}\xspace}
\def\cpizpiz {\ensuremath{C_{\piz\piz}}\xspace}
\def\akpi {\ensuremath{\mathcal{A}_{K\pi}}\xspace}

\def\delalph{\ensuremath{\Delta \alpha_{\pi\pi}}\xspace}

\usepackage{units}
\usepackage{multirow}
\newcommand{\mb}{\ensuremath{m_{B}}}
\newcommand{\mmiss}{\ensuremath{m_\text{miss}}}
\newcommand{\costhetacms}{\ensuremath{\cos\thetacms}}
\newcommand{\thetacms}{\ensuremath{\theta_{B}^*}}
\newcommand{\Bztokspiz} {\ensuremath{\Bz \to \KS\piz}}

\def\cf {\ensuremath{C_f}} 
\def\sf{\ensuremath{S_f}} 

\long\def\inst#1{\par\nobreak\kern 4pt\nobreak
    {\it #1}\par\vskip 10pt plus 3pt minus 3pt}

\begin{document}
{\pagestyle{empty}

\begin{flushright}
\babar-CONF-\BABARPubYear/\BABARConfNumber \\
SLAC-PUB-\SLACPubNumber \\ 
arXiv:\LANLNumber\ [hep-ex]\\
\end{flushright}

\par\vskip 1.2cm

\begin{center}

\Large \bf {\boldmath Measurement of \CP Asymmetries and Branching Fractions in \\ 
$\Bz \to \pip\pim$, $\Bz \to \Kp \pim$, $\Bz \to \piz\piz$, $\Bz \to \Kz \piz$ \\
and Isospin Analysis of $\B \to \pi\pi$ Decays}

\end{center}
\par\vskip 1.2cm

\begin{center}
\large The \babar\ Collaboration\\
\mbox{ }\\
July 31, 2008
\end{center}
\vfill

\begin{center}
\large \bf Abstract
\end{center}
We present preliminary results of improved measurements of the 
\CP-violating asymmetries and branching fractions in the decays 
$\Bz \to \pip\pim$, $\Bz \to \Kp \pim$, $\Bz \to \piz\piz$, and $\Bz \to \Kz \piz$.
This update includes all data taken at the \FourS 
resonance by the \babar\ experiment at the asymmetric \pep2\ $B$-meson factory
at SLAC, corresponding to $467 \pm 5$ million \BB pairs. 
We find
\begin{align*}
   \spipi & =   -0.68 \pm 0.10 \pm 0.03, \\
   \cpipi & =   -0.25 \pm 0.08 \pm 0.02, \\
   {\cal A}_{K\pi} & = -0.107 \pm 0.016 ^{+0.006}_{-0.004}, \\
   \cpizpiz & =  -0.43 \pm 0.26 \pm 0.05, \\
   \BR(\Bztopizpiz) & = ( 1.83 \pm 0.21 \pm 0.13 ) \times 10^{-6}, \\
   \BR(\Bztokzpiz) & = ( 10.1 \pm 0.6 \pm 0.4 ) \times 10^{-6},
\end{align*}
where the first error is statistical and the second is systematic.
We observe \CP violation with a significance of 
$6.7\sigma$ in $\Bz\to\pip\pim$ and
$6.1\sigma$ in $\Bz\to\Kp\pim$.
Constraints on the Unitarity Triangle angle $\alpha$ are determined 
from the isospin relation between all $\B \to \pi\pi$ rates and 
asymmetries. 

\vfill
\vspace{1.5cm}
\begin{center}

Submitted to the 34$^{\rm th}$ International Conference on High-Energy Physics, ICHEP 08,\\
29 July---5 August 2008, Philadelphia, Pennsylvania, USA.

\end{center}

\vspace{0.5cm}
\begin{center}
{\em Stanford Linear Accelerator Center, Stanford University, 
Stanford, CA 94309} \\ \vspace{0.1cm}\hrule\vspace{0.1cm}
Work supported in part by Department of Energy contract DE-AC03-76SF00515.
\end{center}

\newpage
} 

\begin{center}
\small

The \babar\ Collaboration,
\bigskip

%
B.~Aubert,
M.~Bona,
Y.~Karyotakis,
J.~P.~Lees,
V.~Poireau,
E.~Prencipe,
X.~Prudent,
V.~Tisserand
\inst{Laboratoire de Physique des Particules, IN2P3/CNRS et Universit\'e de Savoie, F-74941 Annecy-Le-Vieux, France }
J.~Garra~Tico,
E.~Grauges
\inst{Universitat de Barcelona, Facultat de Fisica, Departament ECM, E-08028 Barcelona, Spain }
% ...

\end{center}\newpage

\section{INTRODUCTION}
\label{sec:Introduction}

Large \CP-violating effects~\cite{largeCPV} in the $B$-meson system
are among the most remarkable predictions of the Cabibbo--Kobayashi--Maskawa (CKM) 
quark-mixing model~\cite{Ckm}. 
These predictions have been confirmed in recent years by the \babar\
and Belle collaborations, both in the interference of \Bz decays to
\CP\ eigenstates with and without $\Bz$--$\Bzb$
mixing~\cite{CPV_beta,BellePRL2007,BaBarPRL2007} 
and directly, in the interference between the
decay amplitudes~\cite{refConj} in $\Bz\to\Kp\pim$~\cite{BaBarPRL2007,BelleNature2008}.  

Effective constraints on physics beyond the Standard Model (SM) are 
provided by high-precision measurements of quantities whose 
SM predictions suffer only small theoretical uncertainties. 
Both experimental and theoretical uncertainties often partially
cancel out in the determination of \CP-violating asymmetries, 
which makes \CP-violation measurements a sensitive probe for effects 
of yet-undiscovered additional interactions and heavy particles 
that are introduced by extensions to the SM.  
All measurements of \CP\ violation to date are in agreement with the indirect
predictions from global SM fits~\cite{ref:CKMfitter,UTfit} that are based on
measurements of the magnitudes of the elements of the
CKM quark-mixing matrix;
this strongly constrains~\cite{CKMnewphys} the flavor
structure of SM extensions.

The CKM Unitarity Triangle angle $\alpha \equiv \arg\left[-V_{\rm
td}^{}V_{\rm tb}^{*}/V_{\rm ud}^{}V_{\rm ub}^{*}\right]$ is
measured through
interference between decays with and without 
$\Bz$--$\Bzb$ mixing.
Multiple measurements of $\alpha$, with
different decays, further test the consistency of the CKM model.  The
time-dependent asymmetry in \Bztopippim is proportional to \stwoa in
the limit that only the $b \to u$ (``tree'') quark-level amplitude 
contributes to this decay. 
%
In the presence of $b \to d$ (``penguin'') amplitudes, the time-dependent 
asymmetry in \Bztopippim is modified to
\begin{equation}
\begin{split}\label{eq:asymmetry}
a(\Delta t) = \frac{|\Abar(\deltat)|^{2} - |A(\deltat)|^{2}}{|\Abar(\deltat)|^{2} + |A(\deltat)|^{2}} & 
        = \spipi \sin{(\deltamd\deltat)} -  \cpipi \cos{(\deltamd\deltat)} \\
\cpipi & = \frac{|A|^{2} - |\Abar|^{2}}{|A|^{2} + |\Abar|^{2}} \\
\spipi & = \sqrt{1 - \cpipi^{2}} \sin{(2\alpha - 2\delalph)} = \sqrt{1 - \cpipi^{2}} \sin{2\alpha_{\rm eff}},
\end{split}
\end{equation}
where \deltat is the difference between the proper decay times 
of the signal- and tag-side neutral $B$ mesons and
\deltamd is the \Bz mixing frequency.  Both the phase
difference $\delalph = \alpha - \alpha_{\rm eff}$ and the direct \CP\ asymmetry \cpipi may 
differ from zero due to the penguin contribution to the \Bztopippim 
decay amplitude $A$.

The magnitude and relative phase of the penguin contribution to
the asymmetry \spipi\ may be unraveled with an analysis of
isospin relations between the
\Btopipi decay amplitudes~\cite{Isospin}.  
The amplitudes $A^{ij}$ of the $B\to \pi^i\pi^j$ decays 
and $\Abar^{ij}$ of the $\Bbar \to \pi^i\pi^j$ decays
satisfy the relations
\begin{equation}\label{eq:isospin}
\begin{split}
A^{+0} = \frac{1}{\sqrt{2}}A^{+-} + A^{00},\\
\Abar^{-0} = \frac{1}{\sqrt{2}}\Abar^{+-} + \Abar^{00}.
\end{split}
\end{equation}
The shape of the
corresponding isospin triangle is determined from measurements of the
branching fractions and time-integrated \CP asymmetries for each of the \Btopipi
decays. 
No gluonic penguin amplitudes are present in the $\Delta I = 3/2$ decay \Btopipiz,
so, neglecting electroweak (EW) penguins, $A^{+0} = \Abar^{-0}$.
We define the direct \CP 
asymmetry \cpizpiz in \Bztopizpiz 
as
\begin{equation}
\begin{split}\label{eq:directCP}
  \cpizpiz & = \frac{|A^{00}|^{2} - |\Abar^{00}|^{2}}{|A^{00}|^{2} +
    |\bar{A}^{00}|^{2}}. 
\end{split}
\end{equation}
From the difference in shape of these triangles for the \B
and $\Bbar$ decay amplitudes, a constraint on $\delalph$ can be determined with a four-fold ambiguity.  

The phenomenology of the \Btopipi system has been thoroughly studied in a number
of theoretical frameworks and models~\cite{ref:Models}.  Predictions for
the relative size and phase of the penguin contribution vary
considerably, so increasingly precise measurements will help
distinguish among different theoretical approaches and add to our
understanding of hadronic \B decays.

The measured rates and direct \CP-violating asymmetries 
in $B\to K\pi$ 
decays~\cite{pi0pi0_BaBar,BabarBRPiPi,ref:BaBarK0K0,ref:BaBarK0pi0,ref:BelleKpiData,ref:CLEOdata} 
reveal puzzling features that could indicate significant contributions from 
EW penguins~\cite{ref:KpiPuzzle1,ref:KpiPuzzle2}.  
Various methods have been proposed to isolate the Standard Model contribution 
to this process in order to test for signs of new physics.  
Sum rules derived from $U$-spin symmetry relate the rates and asymmetries for 
the decays $\Bz$ or $\Bp$ to $\Kp\pim$, $\Kp\piz$, $\Kz\piz$, and $\Kz\pip$~\cite{ref:SumRule}, 
while $SU(3)$ symmetry can be used to make predictions for the $K\pi$ system
based on hadronic parameters extracted from the $\pi\pi$
system~\cite{ref:KpiPuzzle1}.

\section{THE \babar\ DETECTOR AND DATA SET}
\label{sec:babar}

The data used in this analysis were collected in 1999--2007 with the
\babar\ detector at the \pep2\ asymmetric-energy \B-meson factory at the
Stanford Linear Accelerator Center. A total of $467 \pm 5$ million \BB pairs
were used.  The preliminary results presented here supersede
the results in prior publications~\cite{BaBarPRL2007,pi0pi0_BaBar,ref:BaBarK0pi0}. 
Roughly 22\% more \BB pairs have been added to the \babar\ data set, and improvements have been 
introduced to the analysis technique, boosting the signal significance. 

In the \babar\ detector~\cite{babar}, charged particles are detected 
and their momenta measured by a combination of a five-layer double-sided silicon vertex tracker (SVT)
and a 40-layer drift chamber (DCH) that covers 92\% of the solid angle
in the \Y4S center-of-mass (c.m.) frame, both operating in a 1.5-T solenoidal magnetic field.
Discrimination between charged pions, kaons, and protons is provided by a combination of an 
internally reflecting ring-imaging Cherenkov detector (DIRC), which covers 84\% of the c.m. solid 
angle in the central region of the \babar\ detector and has a 91\% reconstruction efficiency 
for pions and kaons with momenta above \unit[1.5]{\gevc}, and the 
ionization (\dedx) measurements in the DCH. 
Neutral-cluster (photon) positions and energies are measured
with an electromagnetic calorimeter (EMC) consisting of 6580 CsI(Tl)
crystals.  The photon energy resolution is $\sigma_{E}/E = \left\{2.3
  / E(\gev)^{1/4} \oplus 1.9 \right\} \%$, and the angular resolution
from the interaction point is $\sigma_{\theta} = 3.9^{\rm o}/\sqrt{E(\gev)}$.  

 
\section{ANALYSIS METHOD}
\label{sec:Analysis}

Many elements of the measurements discussed in this paper are common to the 
decay modes $\Bztohh (h= \pi \;{\rm or} \; K)$, \Bztopizpiz, 
and $\Bz \to\KS \piz$.  
The signal \B-meson candidates (\Brec) are
formed by combining two particles, either tracks or \piz or \KS candidates.
The event selection differs for each mode, and is described in detail below. 

The number of \B decays and the corresponding \CP asymmetries are
determined in extended unbinned maximum likelihood (M.L.) fits to 
variables described below. 
The likelihood is given by the expression
\begin{equation}
{\cal L} = \exp{\left(-\sum_{i}^{M} n_i \right)}
\prod_{j}^{N} \left[\sum_{i}^{M} n_i {\cal P}_{i}(\vec{x}_j;\vec{\alpha}_i)\right],
\end{equation}
where the product is over the number of events $N$, the sums are over
the event categories $M$, $n_{i}$ is the coefficient for each category
as described below, and the probability-density function (PDF) $\cal
P$ describes the distribution of the variables $\vec{x}$ in terms of
parameters $\vec{\alpha}$. The PDF functional forms are discussed
in Sec.~\ref{sec:hhMethod}, \ref{sec:pizpizMethod}, and \ref{sec:kspizMethod}.

\subsection{\boldmath Track and \KS Selection}
\label{sec:TrackSelection}

For particle identification in the \Bztohh sample, we make use of the 
track's Cherenkov radiation in the DIRC as well as its ionization energy
loss \dedx in the DCH. 

For the DIRC information to be used, we require that each track have 
the associated Cherenkov angle (\thetac) 
measured with at least six signal photons detected in the DIRC, where the value of
\thetac is required to be within 4.0 standard deviations from either 
the pion or kaon hypothesis, which effectively removes any candidate
containing high-momentum protons. 
Electrons are explicitly removed
based primarily on a comparison of the track momentum and the associated energy 
deposition in the EMC, with additional 
information provided by DCH \dedx\ and DIRC \thetac measurements. 
 
The ionization energy loss \dedx in the DCH is used either in combination 
with DIRC information or alone, which enables a 35\% increase in the 
\Bztohh reconstruction efficiency compared to using only the tracks with good 
DIRC information.  
A detailed DCH \dedx\ calibration that we developed for the \Bztohh analysis 
takes into account variations in the mean value and resolution 
of \dedx\ values with respect to changes in the DCH running conditions over time and the 
track's charge, polar and azimuthal angles, and number of ionization samples.  The 
calibration is performed with large high-purity samples ($> 10^6$ events) of protons from 
$\Lambda\to\proton\pim$, pions and kaons from $D^{*+}\to D^0\pi^+\,(\Dz\to\Km\pip)$, 
and $\KS \to\pip\pim$ decays that occur in the vicinity of the interaction region. 

$\KS\to\pip\pim$ candidates are reconstructed from pairs of oppositely
charged tracks. The two-track combinations are required to form a vertex with a
$\chi^2$ probability greater than $0.001$ and a $\pip\pim$ invariant
mass within \unit[$11.2$]{\mevcc} ($3.7\sigma$) of the \KS\
mass~\cite{pdg}.  

\subsection{\boldmath \piz Selection}
\label{sec:pi0Selection}

We form $\piz\to\gamma\gamma$ candidates from pairs of clusters in the EMC 
that are isolated from any charged tracks.
Clusters are required to have a transverse energy deposition consistent with that of a 
photon and to have an energy \unit[$E_{\gamma} > 30$]{\mev} for \Bztopizpiz and 
\unit[$E_{\gamma} > 50$]{\mev} for \Bztokspiz{}. 
We use \piz candidates that fall within the invariant-mass range 
\unit[$110<m_{\gamma\gamma}<160$]{\mevcc}. 

For the \Bztopizpiz sample, we also use \piz candidates from a single
EMC cluster containing two adjacent photons (a merged \piz), or one EMC
cluster and two tracks from a photon conversion to an \epem pair
inside the detector.  To reduce the background from random photon
combinations, the angle $\theta_{\gamma}$ between the photon momentum
vector in the \piz rest frame and the \piz momentum vector in the
laboratory frame is required to satisfy $|\cos{\theta_{\gamma}}| <
0.95$. The \piz\ candidates are fitted kinematically with their mass
constrained to the nominal \piz mass~\cite{pdg}.

Photon conversions are selected from pairs of oppositely
charged tracks with an invariant mass below \unit[30]{\mevcc} whose
combined momentum vector points straight away from the beam spot.  The conversion point
is required to lie inside the detector material. 
Converted photons are combined with photons from single EMC clusters to form
\piz candidates. 

Single EMC clusters containing two photons are selected
with the transverse second moment, $S = \sum_{i} E_{i} \times (\Delta
\alpha_i)^{2}/ E$, where $E_{i}$ is the energy in each CsI(Tl) crystal and
$\Delta \alpha_{i}$ is the angle between the cluster centroid and
the crystal.  The second moment is used to distinguish merged \piz
candidates from both single photons and neutral hadrons.

\subsection{\boldmath Event Selection in \Bztopippim, \Bztokpi, and \Bztopizpiz}
\label{sec:EventSelection}
Two kinematic variables are used in the \Bztohh
and \Bztopizpiz analyses to separate \B-meson decays from the large
$\epem \to \qq \;(q=u,\,d,\,s,\,c)$ combinatoric background~\cite{babar}. 
One is the beam-energy--substituted mass $\mes = \sqrt{ (s/2 + {\bf
    p}_{i}\cdot{\bf p}_{B})^{2}/E_{i}^{2}- {\bf p}^{2}_{B}}$, where
$\sqrt{s}$ is the total \epem c.m.\ energy, $(E_{i},{\bf
  p}_{i})$ is the four-momentum of the initial \epem system in the
laboratory frame, and ${\bf p}_{B}$ is the laboratory momentum of the \B 
candidate.
The other is 
\de $ = E^{*}_{B} - \sqrt{s}/2$, where $E^{*}_{B}$ is
the \B candidate's energy in the c.m.\ frame.

Two additional quantities take advantage of the event topology to
further separate \B decays from the \qq background.  The absolute value of the 
cosine of the
angle $\theta_{\scriptscriptstyle S}$ between the sphericity axes~\cite{Bjorken:1969wi} of
the \B candidate's decay products and that of the remaining tracks and
neutral clusters in the event, computed in the c.m.\ frame, is peaked at 1.0 for 
the jet-like
\qq events but has a flat distribution for \B decays.  We require
$\cossph < 0.7$ for \Bztopizpiz and 
$\cossph < 0.91$ for \Bztohh.  For
the $\Bztohh$ sample, we further require that the second
Fox--Wolfram moment~\cite{R2all} satisfy $R_2 <0.7$ to remove a small remaining
background from $\epem \to \tautau$ events.

To improve the discrimination
against \qq events, a Fisher discriminant \fish is formed as a linear
combination of the sums $L_0\equiv\sum_i |{\bf p}^*_i|$ and
$L_2\equiv\sum_i |{\bf p}^*_i| \cos^2 \theta^*_i$, where
${\bf p}^*_i$ are the momenta and $\theta^*_i$ are the
angles with respect to the thrust axis~\cite{ref:thrust} of the \B candidate, 
both in the c.m.\ frame, of all tracks and
clusters not used to reconstruct the signal \B-meson candidate.
In the case of \Bztopizpiz, we improve the sensitivity of the signal by 
combining \fish with three other quantities in a neural network.
These are the $|\cos \theta_{\scriptscriptstyle S}|$ described above, 
$|\cos\theta_B|$, where $\theta_B$ is the angle between the
center-of-mass momentum vector of the signal $B$ and the beam axis,
and $|\cos\theta_{\scriptstyle T}|$, where $\theta_{\scriptstyle T}$ is
the angle between the thrust axis of the signal $B$-meson's daughters
and the beam axis.

\subsubsection{\boldmath \Bztopippim and \Bztokpi}
\label{sec:hhMethod}

We reconstruct candidate decays
$B_{\rm rec} \to h^+ h^{\prime-}$  from pairs of oppositely charged tracks 
in the polar-angle range $0.35 < \theta_{\rm lab} <2.40 $ that are consistent with originating 
from a common decay point with a $\chi^2$ probability of at least 0.001.
The remaining particles are examined to infer whether 
the other $B$ meson in the event (\Btag) decayed as a $\Bz$ or $\Bzb$ (flavor tag).
We perform an unbinned extended M.L.\ fit 
to separate $\Bztopippim$ and $\Bztokpi$ decays
and determine simultaneously their 
\CP-violating asymmetries \spipi, \cpipi, and \akpi\ and 
the signal and background yields and PDF parameters.  
The fit uses particle-identification, kinematic, event-shape, $B_{\rm tag}$ flavor, and 
$\deltat$ information.

The variables \mes and \de are calculated assuming that both tracks are
charged pions.  The $\Bz \to \pip \pim$ events are described by a Gaussian
distribution for both \mes and \de, where the resolutions are found to be \unit[2.6]{\mevcc}
and \unit[29]{\mev}, respectively.  For each kaon in the final state, the
$\de$ peak position is shifted from zero by an amount that depends on the
kaon momentum, with an average shift of \unit[$-45$]{\mev}.  We require 
\unit[$\mes > 5.20$]{\gevcc} 
and \unit[$\left|\de\right|<0.150$]{\gev}.  The large region
below the signal in $\mes$ effectively determines the background
shape parameters, while the wide range in $\de$ allows us to separate
\Bz decays to all four final states ($\pip\pim$, $\Kp\pim$, $\pip\Km$, and $\Kp\Km$) 
in a single fit.

\begin{figure}[!tbp]
\begin{center}
\includegraphics[width=0.65\linewidth,clip=true]{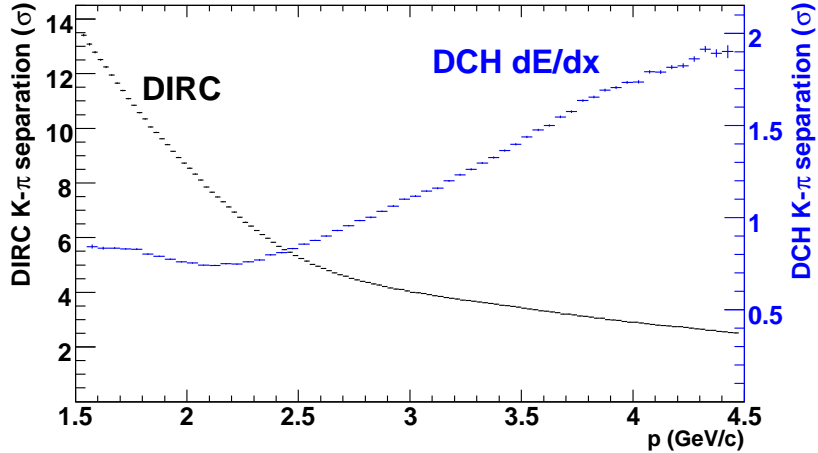}
\caption{\small The average difference between the expected values of DIRC $\theta_{\rm C}$ 
and DCH \dedx\ 
for pions and kaons at $0.35 < \theta_{\rm lab} <2.40 $, divided by the uncertainty, 
as a function of laboratory momentum in $\Bz \to \Kp\pim$ decays in $\babar$.}
\label{fig:DIRC_dEdx}
\end{center}
\end{figure}

We construct $\theta_{\rm C}$ PDFs
for the pion and kaon hypotheses, and \dedx\ PDFs for the 
pion, kaon, and proton hypotheses, separately for each charge. 
The $K$--$\pi$ separations provided by $\theta_{\rm C}$ and \dedx\ are complementary:  
for $\theta_{\rm C}$, the separation varies from $2.5 \sigma$ 
at \unit[4.5]{\gevc} to $13 \sigma$ at 
\unit[1.5]{\gevc}, while for \dedx\ it varies from less than 
$1.0\sigma$ at \unit[1.5]{\gevc} to $1.9 \sigma$ at \unit[4.5]{\gevc} 
(Fig.~\ref{fig:DIRC_dEdx}). For more details, please see Ref.~\cite{BaBarPRL2007}.

We use a multivariate technique~\cite{ref:sin2betaPRL04} to determine the
flavor of $\Btag$.  Separate neural networks are
trained to identify leptons and kaons from \B\ and $D$ decays
and soft pions from $D^*$ decays.  Events
are assigned to one of seven mutually exclusive tagging categories
(including untagged events) based on the estimated average mistag
probability and the source of the tagging information.  
The quality of tagging is expressed in
terms of the effective efficiency $Q = \sum_k \epsilon_k(1-2w_k)^2$,
where $\epsilon_k$ and $w_k$ are the efficiencies and mistag
probabilities, respectively, for events tagged in category $k$. The
difference in mistag probabilities is given by $\Delta w = w_{\Bz} -
w_{\Bzb}$.  Table~\ref{tab:tagging} summarizes the tagging performance
measured in a large data sample of fully reconstructed neutral \Bflav\ decays to
$D^{(*)-}(\pip,\, \rho^+,\, a_1^+)$.

\begin{table}[!tbp]
\caption{\small Average tagging efficiency $\epsilon$, average mistag fraction $w$,
mistag fraction difference $\Delta w = w(\Bz) - w(\Bzb)$, and effective tagging efficiency 
$Q$ for signal events in each tagging category.  The quantities are measured in the 
large-statistics \Bflav\ sample of fully reconstructed neutral \B-meson decays.}

\begin{center}
\begin{tabular}{cr@{~}c@{~}lr@{~}c@{~}lr@{~}c@{~}lr@{~}c@{~}l} \hline\hline
Category & \multicolumn{3}{c}{$\epsilon\,(\%)$} & \multicolumn{3}{c}{$w\,(\%)$} & \multicolumn{3}{c}{$\Delta w\,(\%)$} &
\multicolumn{3}{c}{$Q\,(\%)$} \rule[-2mm]{0mm}{6mm} \\\hline
{\tt Lepton}    & $8.96  $&$ \pm $&$ 0.07  $&$ 2.9  $&$ \pm $&$ 0.3 $&$  0.2 $&$ \pm $&$ 0.5 $&$ 7.95  $&$ \pm $&$ 0.11$\\
{\tt Kaon\,I}   & $10.81 $&$ \pm $&$ 0.07  $&$ 5.3  $&$ \pm $&$ 0.3 $&$  0.0 $&$ \pm $&$ 0.6 $&$ 8.64  $&$ \pm $&$ 0.14$\\
{\tt Kaon\,II}  & $17.18 $&$ \pm $&$ 0.09  $&$ 14.5 $&$ \pm $&$ 0.3 $&$  0.4 $&$ \pm $&$ 0.6 $&$ 8.64  $&$ \pm $&$ 0.17$\\
{\tt Kaon\,Pion}& $13.67 $&$ \pm $&$ 0.08  $&$ 23.3 $&$ \pm $&$ 0.4 $&$ -0.6 $&$ \pm $&$ 0.7 $&$ 3.91  $&$ \pm $&$ 0.12$\\
{\tt Pion}      & $14.19 $&$ \pm $&$ 0.08  $&$ 32.6 $&$ \pm $&$ 0.4 $&$  5.1 $&$ \pm $&$ 0.7 $&$ 1.73  $&$ \pm $&$ 0.09$\\
{\tt Other}     & $9.55  $&$ \pm $&$ 0.07  $&$ 41.5 $&$ \pm $&$ 0.5 $&$  3.8 $&$ \pm $&$ 0.8 $&$ 0.28  $&$ \pm $&$ 0.04$\\
\hline
Total           &        &       &       &        &       &       &        &       &       &$ 31.1 $&$ \pm $&$ 0.3$ \rule[-2mm]{0mm}{6mm} \\\hline\hline
\end{tabular}
\end{center}
\label{tab:tagging}
\end{table}

The time difference $\deltat = \Delta z/\beta\gamma c$ is obtained
from the known boost of the $\epem$ system ($\beta\gamma = 0.56$) and
the measured distance $\Delta z$ along the beam ($z$) axis between the
\Brec\ and \Btag\ decay vertices.  
A description of the inclusive reconstruction of the
\Btag{} vertex is given in \cite{ref:BaBarsin2beta}.
We require
\unit[$\left|\deltat\right|<20$]{\ps} and \unit[$\sigma_{\deltat} < 2.5$]{\ps}, where
$\sigma_{\deltat}$ is the error on $\deltat$ determined separately for
each event. The signal \deltat PDF for $\Bztopippim$ is given by
\begin{equation}
\begin{split}
f^{\pm}_{k}(\deltat_{\rm meas}) = \frac{e^{-|\deltat|/\tau}}{4\tau}
           \Bigl\{ &( 1 \mp \Delta w )  \\
                   & \pm ( 1-2w_{k} ) \bigl[  \spipi \sin{(\deltamd\deltat)} -  \cpipi\cos{(\deltamd\deltat)}    
                   \bigr]   
           \Bigr\} \otimes R(\deltat_{\rm meas} - \deltat),
\end{split}
\end{equation} 
where $f^{+}_{k}$ ($f^{-}_{k}$) indicates a \Bz (\Bzb) flavor tag and
the index $k$ indicates the tagging category.  The resolution function
$R(\deltat_{\rm meas} - \deltat)$ for signal candidates is a sum of three 
Gaussian functions,
identical to the one described in Ref.~\cite{ref:BaBarsin2beta}, with
parameters determined from a fit to the \Bflav\ sample (including
events in all seven tagging categories).  The background $\deltat$
distribution is also modeled as the sum of three Gaussians, where
the common parameters used to describe the background shape for all
tagging categories are determined simultaneously with the \CP
parameters in the maximum likelihood fit.

The M.L.\ fit includes 28 components: \Bz\ signal decays and background with
the final states $\pip\pim$, $\Kp\pim$, $\Km\pip$, and $\Kp\Km$
where either the positively charged or the negatively charged track,
or both, have good DIRC information ($2 \times 4 \times 3 = 24$ 
components) plus the $\proton\pim$, $\proton\Km$, $\pip\antiproton$
and $\Kp\antiproton$ background components where the (anti)proton
has no DIRC information. The $\Kpm\pimp$ event yields are parameterized as
$n_{\Kpm\pimp}=n_{K\pi}\left(1\mp \akpi^{\rm raw} \right)/2$. All other coefficients are 
products of the fraction of events in each tagging category, taken from
\Bflav\ events, and the event yield.  The background PDFs
are a threshold function~\cite{ref:argus}
for \mes and a second-order polynomial for \de.  The \fish
PDF is a sum of two asymmetric Gaussians for both the signal and background.  
We used large samples of simulated $B$ decays to investigate the effects of 
backgrounds from other $B$ decays on the determination of the \CP-violating
asymmetries in \Bztopippim and \Bztokpi and determined them to be negligible.

\subsubsection{\boldmath \Bztopizpiz}
\label{sec:pizpizMethod}

\Bztopizpiz events are identified with an M.L.\ fit to the variables
\mes, \de, and \emph{NN}, the output of the event-shape neural network. 
We require \unit[$\mes>5.20$]{\gevcc} and
\unit[$|\de| < 0.2$]{\gev}.  Tails in the EMC response produce a
correlation between \mes and \de, so a two-dimensional PDF, derived
from detailed Monte Carlo (MC) simulation, is used to describe
signal. The \emph{NN} distribution is binned in ten bins (equally populated
for signal) and described by a parametric step-function PDF with 9 height
parameters taken from the MC and fixed in the fit. \Bflav\ data
are used to verify that the MC accurately reproduces the \emph{NN}
distribution.  The \qq background PDFs are a threshold function~\cite{ref:argus} for
\mes, a second-order polynomial for \de, and a parametric step function for \emph{NN}.
For \qq events, \emph{NN} is not distributed uniformly across the bins but rises
sharply toward the highest bins.  We see a small linear correlation
between the shape parameter of the \mes threshold function and the \emph{NN}
bin number, and this linear relation is taken into account in the
fit. All \qq background PDF parameters are allowed to float in the
M.L.\ fit.

The decays \Bptorhoppiz and $\Bztokspiz\ (\KS\to\piz\piz)$ add $71\pm
10$ background events to \Bztopizpiz and are included as an
additional fixed component in the M.L.\ fit.  We model these \B-decay
backgrounds with a two-dimensional PDF to describe \mes and \de, and
with a step function for \emph{NN}, all taken from MC simulation.

The time-integrated \CP asymmetry is measured by the \B-flavor
tagging algorithm described previously.  The fraction of events in each tagging
category is also constrained to the corresponding fraction determined from
MC simulation.  The PDF coefficient for the \Bztopizpiz signal is
given by the expression
\begin{equation}
n_{\piz\piz, k} = \frac{1}{2} f_{k} N_{\piz\piz} \Bigl\{ 1 - s_j
  (1-2\chi)(1-2w_k) \cpizpiz \Bigr\},
\end{equation}
where $f_k$ is the fraction of events in the tagging category $k$,
$N_{\piz\piz}$ is the number of \Bztopizpiz decays, $\chi = 0.188 \pm
0.003$~\cite{pdg} is the time-integrated \Bz mixing probability, and $s_j=+1(-1)$
when the \Btag\ is a \Bz (\Bzb).

\subsection{\boldmath\Bztokzpiz}
\label{sec:kspizMethod}

For each \Bztokspiz candidate, two independent kinematic variables are
computed.  The first one is $\mb$, the invariant mass of the
reconstructed $B$ meson, \Brec.  The second one is $\mmiss$, the
invariant mass of the other $B$, \Btag, computed from the known beam
energy, by applying a mass constraint to \Brec~\cite{ref:kspi0prd05}.  For
signal decays, $\mb$ and $\mmiss$ peak near the \Bz mass with 
resolutions of \unit[$\sim 36$]{\mevcc} and \unit[$\sim5.3$]{\mevcc}, respectively.
Both the $\mmiss$ and $\mb$ distributions exhibit a low-side tail due to 
the leakage of energy deposits out of the EMC.  We select candidates
within the ranges \unit[$5.11<\mmiss<5.31$]{\gevcc} and
\unit[$5.13<\mb<5.43$]{\gevcc}, which include a signal peak and a
``sideband'' region for background characterization. In
the events with more than one reconstructed
candidate (0.8\% of the total), we select the candidate with the smallest
$\chi^2=\sum_{i=\piz,\KS} (m_i-m'_i)^2/\sigma^2_{m_i}$, where $m_i$
($m'_i$) is the measured (nominal) mass and $\sigma_{m_i}$ is the
estimated uncertainty on the measured mass of particle $i$.

We exploit topological observables, computed in the c.m.\ frame, to
discriminate jet-like $\epem \to \qqbar$ events $(q=u,d,s,c)$ from the
nearly spherical \BB{} events.  
In order to reduce the number of background events, we require $L_2/L_0<0.55$, 
where $L_j\equiv\sum_i |{\bf p}^*_i| \cos^j \theta^*_i$ and 
$\theta^*_i$ are computed with respect to the sphericity
axis~\cite{Bjorken:1969wi} of the \Brec\ candidate.
We compute
$\costhetacms$, the cosine of the angle between the direction of the
$B$ meson and the nominal direction of the magnetic field ($z$ axis).
This variable is distributed as $1-\cos^2\thetacms$ for signal events
and is nearly flat for background events. We select events with
$|\costhetacms| < 0.9$. We also use the distributions of $L_2/L_0$ and
$\costhetacms$ to discriminate the signal from the residual background
in a M.L.\ fit.  Using a full detector simulation, we estimate that
our selection retains $(34.2 \pm 1.2)\%$ of the signal events, where
the error includes both statistical and systematic contributions.  The
selected sample of \Bztokspiz{} candidates is dominated by random
$\KS\piz$ combinations from $\epem\to\qqbar$
fragmentation.  Using large samples of simulated \BB{} events, we find
that backgrounds from other \B-meson decays are small, ${\cal O}(0.1\%)$;
however, we study in detail the effect of a number of specific $B$
decay channels. The dominant ones are $B^+\ra\rho^+\KS$,
$B^+\ra K^{*+}\piz$, and $B^+\ra\KS\piz\pi^+$, and we include this effect in
our study of the systematic errors.

For the
\Bztokspiz{} decay, where no charged particles are present at the
decay vertex, we compute the decay point of the \Brec{} using the knowledge of the 
\KS{} trajectory from the measurement of $\pi^+ \pi^-$ momenta and the knowledge of
the average interaction point~\cite{ref:kspi0prd08}. 

We extract the signal yield from an extended unbinned M.L.\ fit to 
kinematic, event-shape, flavor-tag, and decay-time quantities. 
The use of tagging and decay-time information in the M.L.\ fit further improves 
discrimination between signal and background.
We have verified that all correlations
are negligible, and so construct the likelihood function as a product of
one-dimensional PDFs.  Residual correlations are taken into account in
the systematic uncertainty, as explained below.

The PDFs for signal events are parameterized based on a large sample of
fully reconstructed $B$ decays in data and from simulated events.  For
background PDFs, we select the functional form from the
background-dominated sideband regions in the data.

The likelihood function is defined as:
{\small\begin{eqnarray}
\label{eq:ml}
\lefteqn{
{\cal L}(\sf,\cf,N_{\rm sig},N_{\rm bkg},f_{\rm sig},f_{\rm bkg},\vec{\alpha}) =\frac{e^{-(N_{\rm sig}+N_{\rm bkg})}}{N\,!}} &&  \\
&& \times \prod_{i \in g}
    \left[ N_{\rm sig} f_{\rm sig} \epsilon^{c}_{\rm sig}{\cal P}_{\rm sig}(\vec{x}_i,\vec{y}_i;\sf,\cf) + 
     N_{\rm bkg} f_{\rm bkg} \epsilon^{c}_{\rm bkg} {\cal P}_{\rm bkg}(\vec{x}_i,\vec{y}_i;\vec{\alpha}) \right]
  \nonumber\\ 
&& \times \prod_{i \in  b}
    \left[ N_{\rm sig} (1-f_{\rm sig}) \epsilon^{c}_{\rm sig} {\cal P}'_{\rm sig}(\vec{x}_i;\cf) + 
     N_{\rm bkg} (1-f_{\rm bkg}) \epsilon_{\rm bkg}^{c} {\cal P}'_{\rm bkg}(\vec{x}_i;\vec{\alpha}) \right], \nonumber
\end{eqnarray}} 

\noindent
where the $N$ selected events are partitioned into two subsets: $i \in
g$ events have 
\deltat
information, while $i \in b$ events do
not. Here, $f_{\rm sig}$ ($f_{\rm bkg}$) is the fraction of signal (background) events $\in
 g$, and $1-f_{\rm sig}$ ($f_{\rm bkg}$) is the fraction of events $\in b$. The
 probabilities ${\cal P}_{\rm sig}$ and ${\cal P}_{\rm bkg}$ are products of PDFs for
 signal (sig) and background (bkg) hypotheses evaluated for the
 measurements
 $\vec{x}_i=\{\mb,\;\mmiss,\;L_{2}/L_{0},\;\costhetacms,\;\text{flavor
 tag},\;\text{tagging category}\}$ and
 $\vec{y}_i=\{\deltat,\sigma_{\deltat}\}$.  
${\cal P}'_{\rm sig}$ and
 ${\cal P}'_{\rm bkg}$ are the corresponding probabilities for events without
 $\deltat$ information.  In the formula, $\vec{\alpha}$ represents the
set of parameters that define the shape of the PDFs.  Along with the
\CP asymmetries \sf\ and \cf, the fit extracts the yields $N_{\rm sig}$ and
$N_{\rm bkg}$, the fraction of events $f_{\rm sig}$ and $f_{\rm bkg}$, and the parameters
$\vec{\alpha}$ that describe the background PDFs.  

\section{RESULTS AND SYSTEMATIC UNCERTAINTIES}
\label{sec:Physics}

\subsection{\boldmath \Bztopizpiz Results}
\label{sec:pizpizResults}

\begin{figure}[!tbp]
\begin{center}
\includegraphics[width=0.46\textwidth]{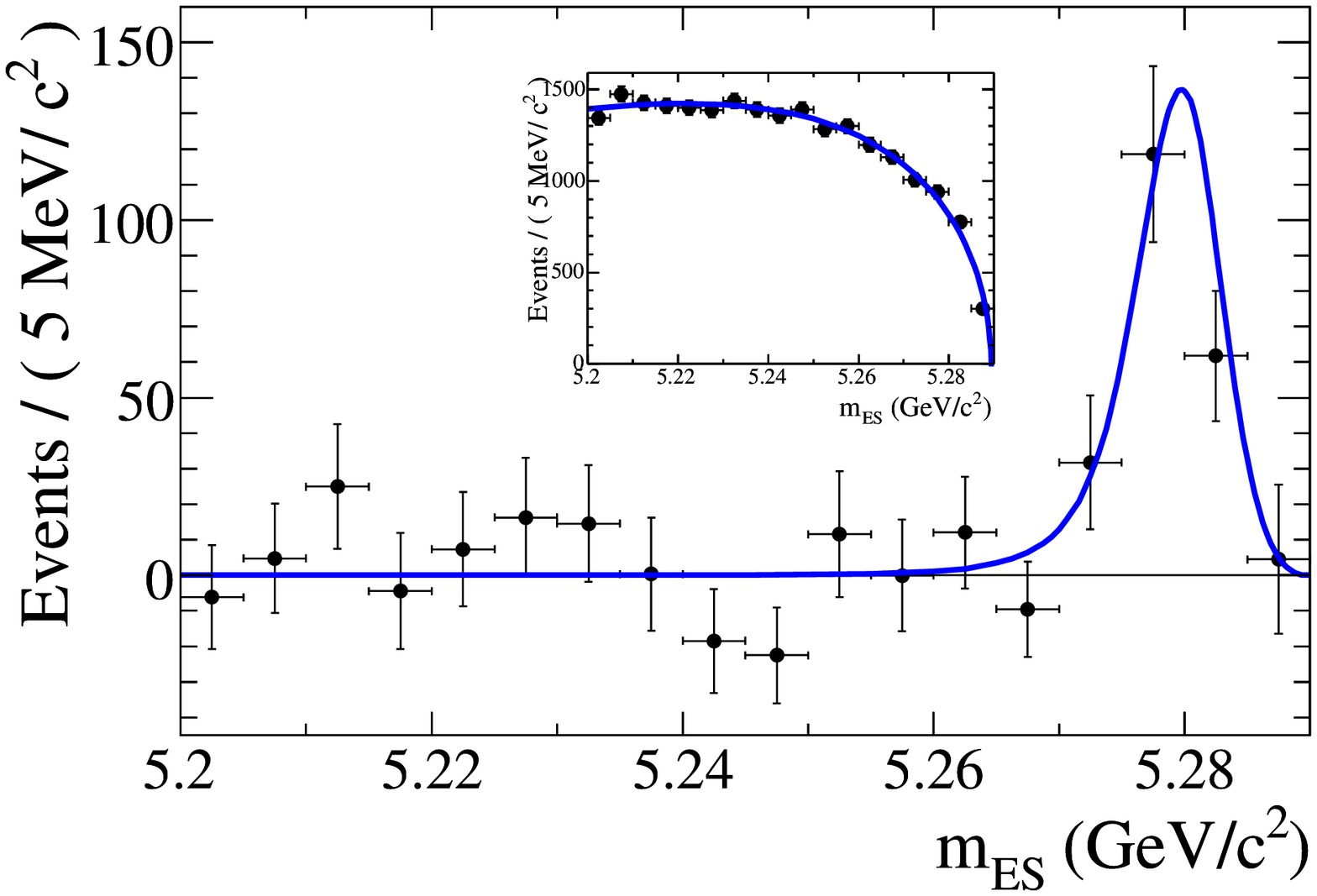}
\includegraphics[width=0.46\textwidth]{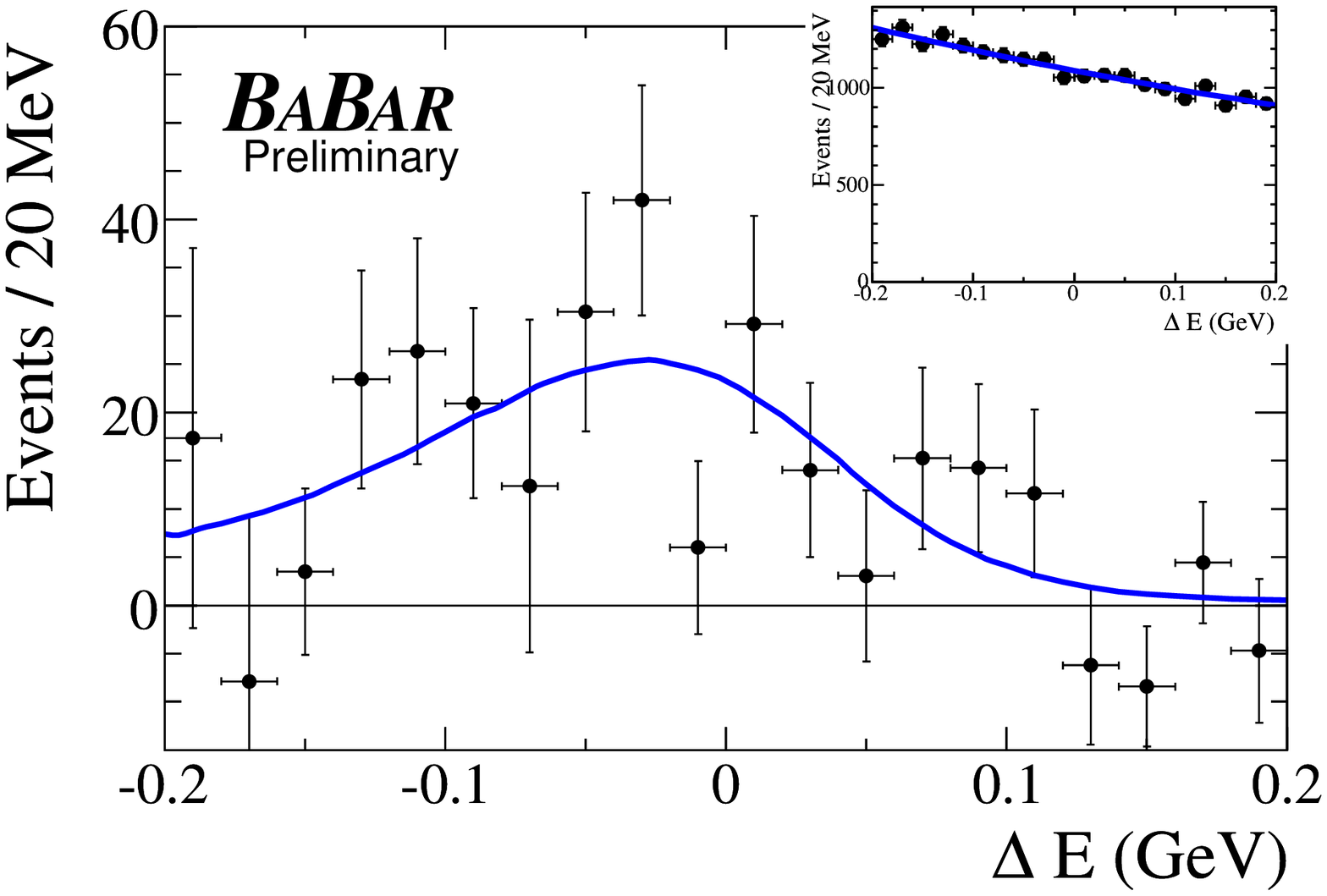}
\includegraphics[width=.46\textwidth]{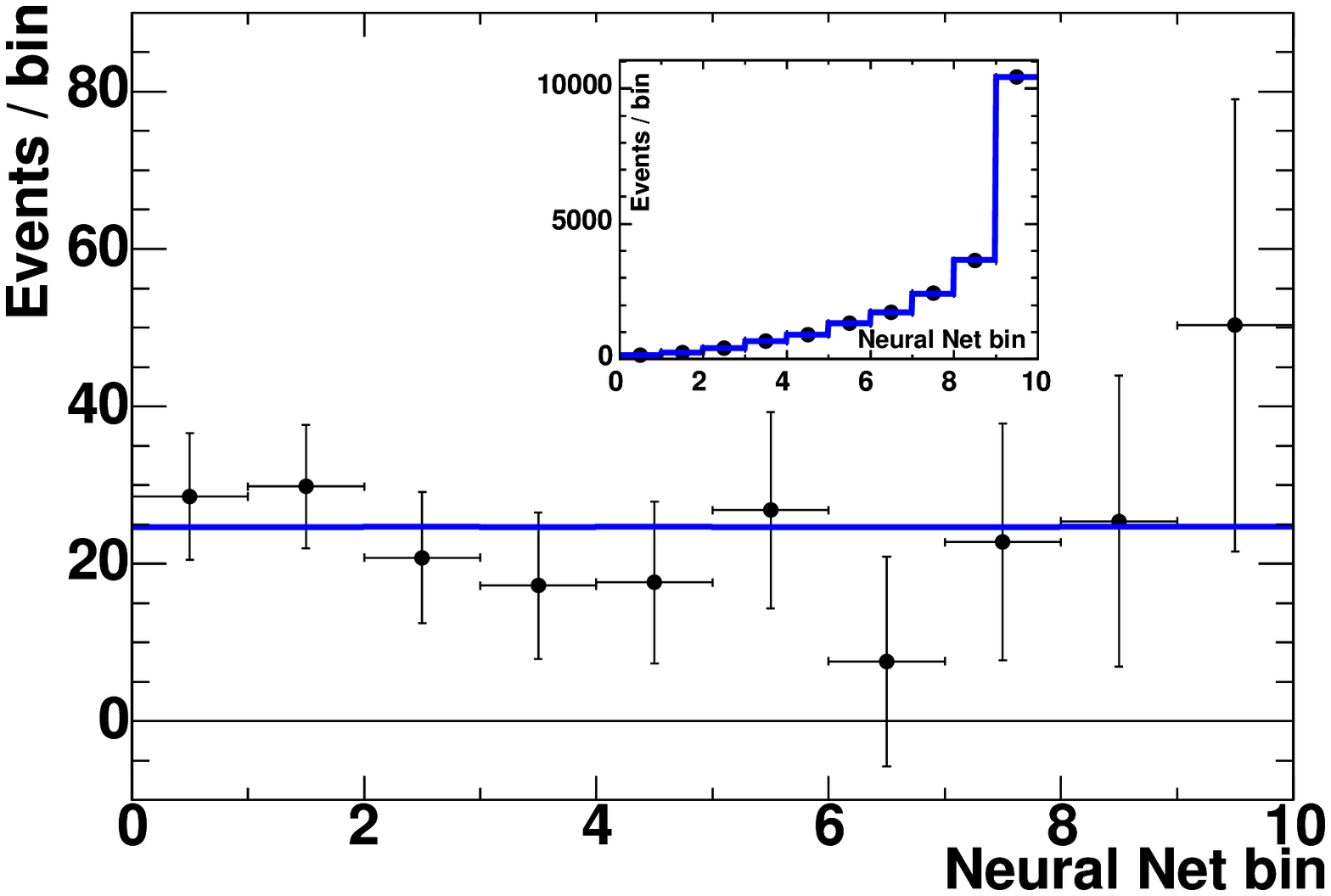}
\caption{ \small \sPlots for \Bztopizpiz signal (background shown in the inset plots):
\emph{(top left)} \mes, 
\emph{(top right)} \de, 
\emph{(bottom)} the binned \emph{NN}. The line
in each plot shows the corresponding PDF.
}
\label{fig:pizpiz}
\end{center}
\end{figure}

Results from the M.L.\ fit for the \Bztopizpiz decay mode are
summarized in Table~\ref{tab:resultsA}. 
Distributions of \mes, \de, and \emph{NN} for \Bztopizpiz
are shown in Fig.~\ref{fig:pizpiz}, where a weighting and background-subtraction 
technique, \sPlots~\cite{ref:splots}, is used to display
the signal events. The same technique is used to display the \qq
background as well, shown in the insets.

\begin{table*}[!bp]
\caption{ \small 
  Results for the \Bztopizpiz and \Bztokzpiz decay modes: signal yields $N_{\rm sig}$, efficiencies, 
  branching fractions, and time-integrated \CP asymmetries. When two uncertainties 
  are given, the first is statistical and the second systematic.}
\label{tab:resultsA}
\begin{center}
\begin{tabular}{ccccc}
\hline\hline
 & $N_{\rm sig}$  & Efficiency       & Branching fraction   & Asymmetry       \\ \hline 
\Bztopizpiz & $247\pm 29$  & $(28.8\pm1.8)\%$ & $( 1.83 \pm 0.21 \pm 0.13 ) \times 10^{-6}$ & $-0.43 \pm 0.26 \pm 0.05$ \\
\Bztokspiz & $556 \pm 32$  & $(34.2\pm1.2)\%$ & $( 10.1 \pm 0.6 \pm 0.4 ) \times 10^{-6}$ & ~\cite{ref:CPVKspi02008} \\
\hline\hline
\end{tabular}
\end{center}
\end{table*}

The uncertainty in the efficiency for the \Bztopizpiz decay mode is
dominated by a $3\%$ systematic uncertainty per \piz, estimated from a
study of $\tau \to \pi\piz\nu_{\tau}$ decays. There is an additional
$1.0\%$ uncertainty in the resolution of the signal shape and a
$0.45\%$ uncertainty due to the limited knowlegde of the \mes and \de
peak positions in data, estimated by shifting the \mes and \de means
and resolutions by amounts determined from MC--data comparison in a
control sample of $\Bp\to \pi^+\pi^0$ events. We also take an
uncertainty of 1.5\%, determined from the \Bflav\ sample, due to the
$\cossph$ requirement. Systematic uncertainties involving the M.L.\
fit are evaluated by varying the PDF parameters and refitting the
data. These contribute an uncertainty of 8.3 events to the
branching-fraction measurement and an uncertainty of 0.05 to
$C_{\piz\piz}$. The various systematics sources are tabulated in
Table~\ref{tab:pizpizsyst}.

\begin{table*}[!hptb]
\small
\caption{ \small
  Systematic uncertainties in the determination of the \Bztopizpiz signal yield ($N_{\piz\piz}$)
  and branching fraction, and the direct \CP asymmetry $C_{\piz\piz}$.
  The total branching-fraction systematic is the sum in quadrature of the uncertainties on 
  the signal yield, the signal efficiency, and the $B$-meson counting. }
\label{tab:pizpizsyst}
\smallskip
\begin{center}
\begin{tabular}{l|c@{\hspace{3em}}c@{\hspace{3em}}c@{\hspace{3em}}}
\hline \hline
Source                &   $N_{\piz\piz}$   & $\sigma_{\rm syst}({\cal B})/{\cal B}$ & $C_{\piz\piz}$ \\ \hline
Peaking background    &   $\pm 4.9$        &          & $\pm 0.030$     \\
Tagging               &   $\pm 0.35$       &          & $\pm 0.034$     \\
Background shape      &   $\pm 5.5$        &          & $\pm 0.023$     \\
Signal shape          &   $\pm 3.8$        &          & $\pm 0.020$     \\ 
\hline
Total fit systematics &   $\pm 8.3$        &  3.4\%  & $\pm 0.055$     \\ 
\hline
&&&\\
\piz efficiency       &  & $6.0\%$ \\
$\cossph$ selection    &  & $1.5\%$ \\
neutrals resolution   &  & $1.0\%$ \\
\mes and \de shape    &  & $0.5\%$ \\ 
\hline
Total efficiency systematics &  & $6.3\%$ \\
\hline
&&&\\
Number of $\BB$ pairs          & & $ 1.1\%$  \\ 
&&&\\
\hline
Total systematic error & & $7.2\%$ & $\pm 0.055$ \\ 
\hline 
\hline
\end{tabular}
\end{center}
\end{table*}


\subsection{\boldmath \Bztopippim and \Bztokpi Results}
\label{sec:hhResults}

\begin{table*}[!tbp]
\caption{ \small 
  Results for the \Bztohh decay modes.  For each
  mode, the number of signal events $N_{\rm sig}$ and \CP asymmetries are shown.  
  Statistical, followed by systematic, uncertainties are given for the asymmetries.}
\label{tab:resultsB}
\begin{center}
\begin{tabular}{l|cc}
\hline\hline
Mode        & $N_{\rm sig}$      & Asymmetry       \\ \hline 
\Bztopippim &  $1394\pm 54$& $\spipi = -0.68 \pm 0.10 \pm 0.03;  \;\; \cpipi = -0.25 \pm 0.08 \pm 0.02$ \\
\Bztokpi    &  $5410\pm 91$ & $\akpi  = -0.107 \pm 0.016^{+0.006}_{-0.004}$ \\
\hline\hline
\end{tabular}
\end{center}
\end{table*}

\begin{figure}[!tbp]
\begin{center}

  \includegraphics[width=0.32\linewidth]{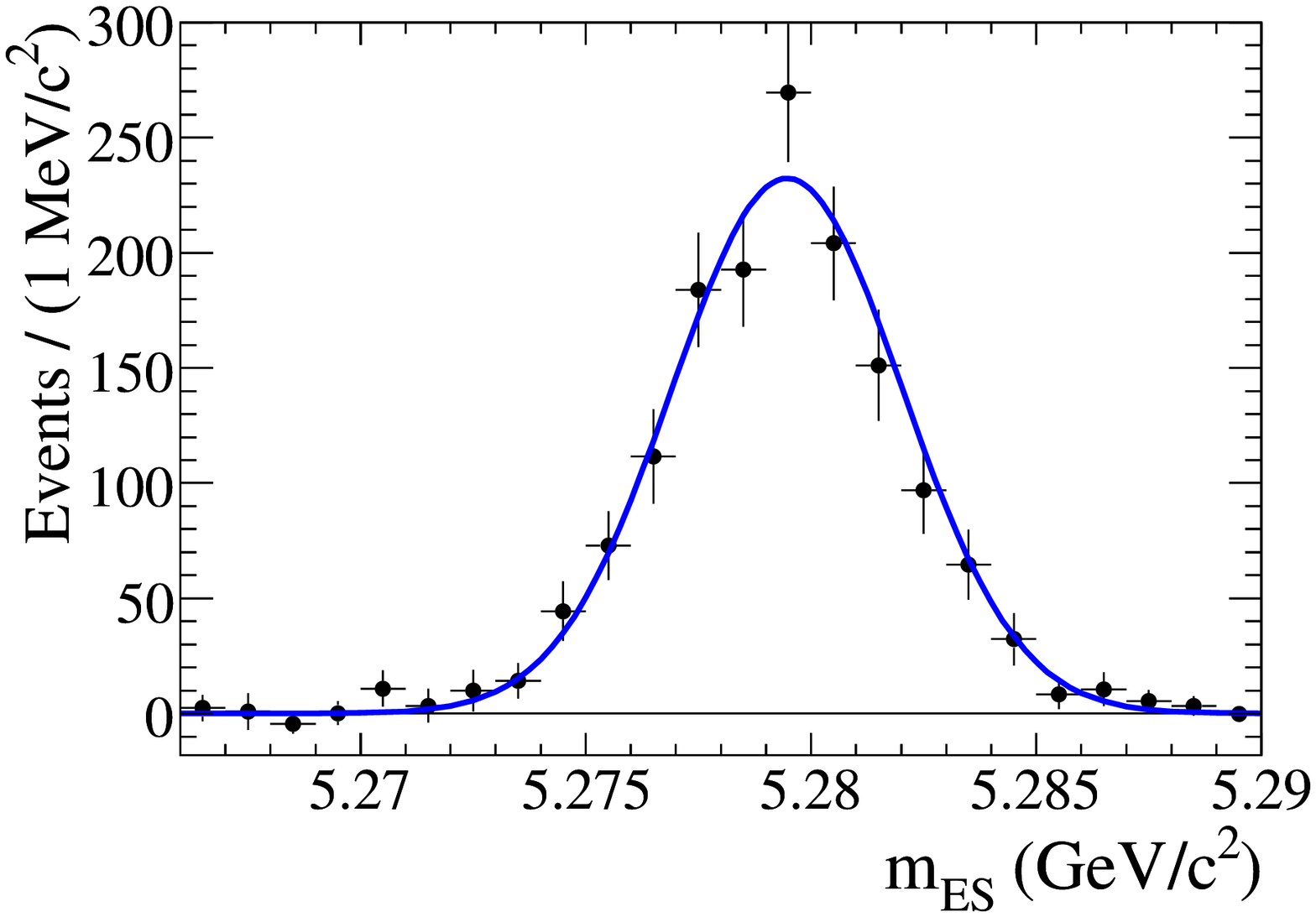}~~
  \includegraphics[width=0.31\linewidth]{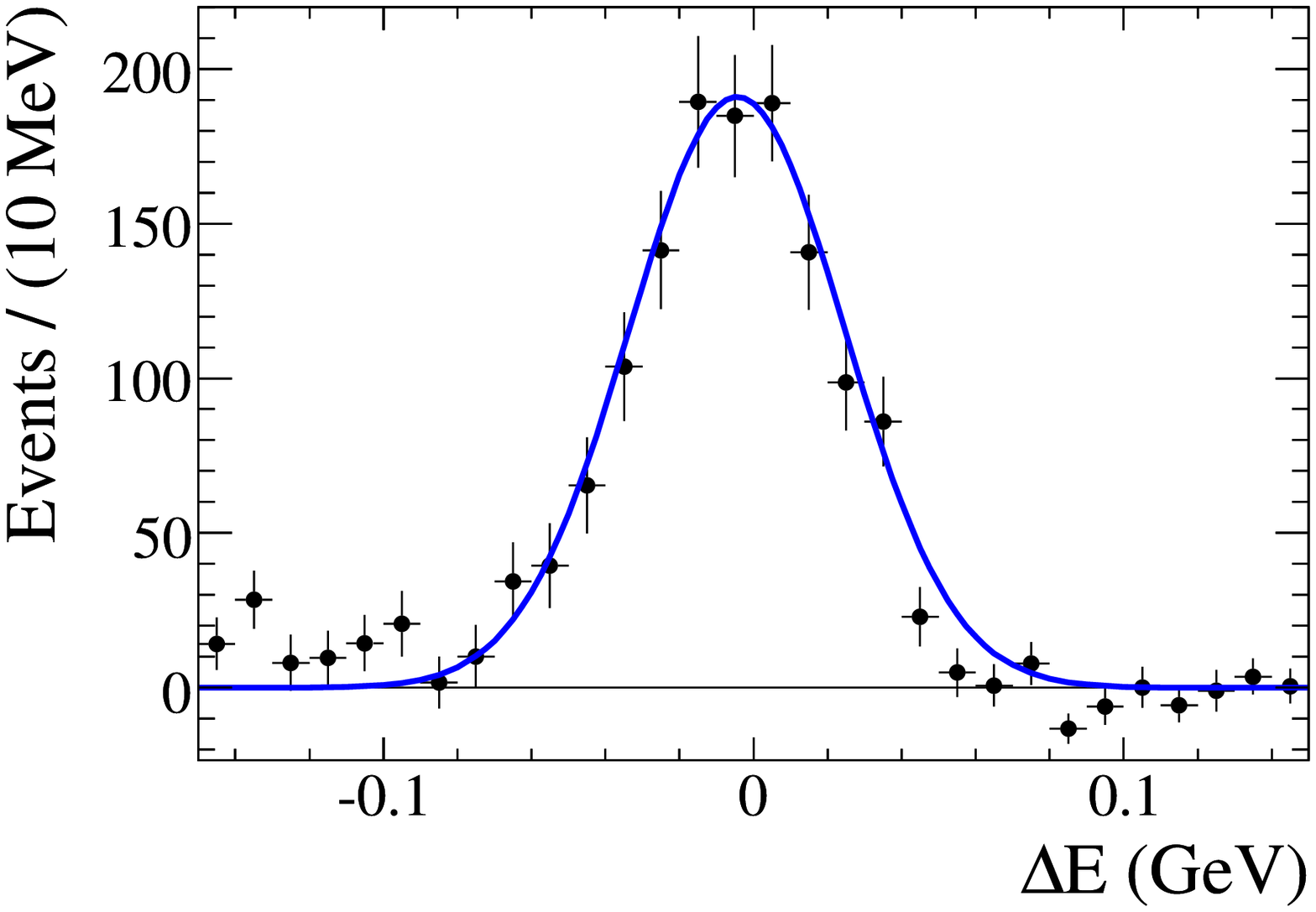}~~
  \includegraphics[width=0.31\linewidth]{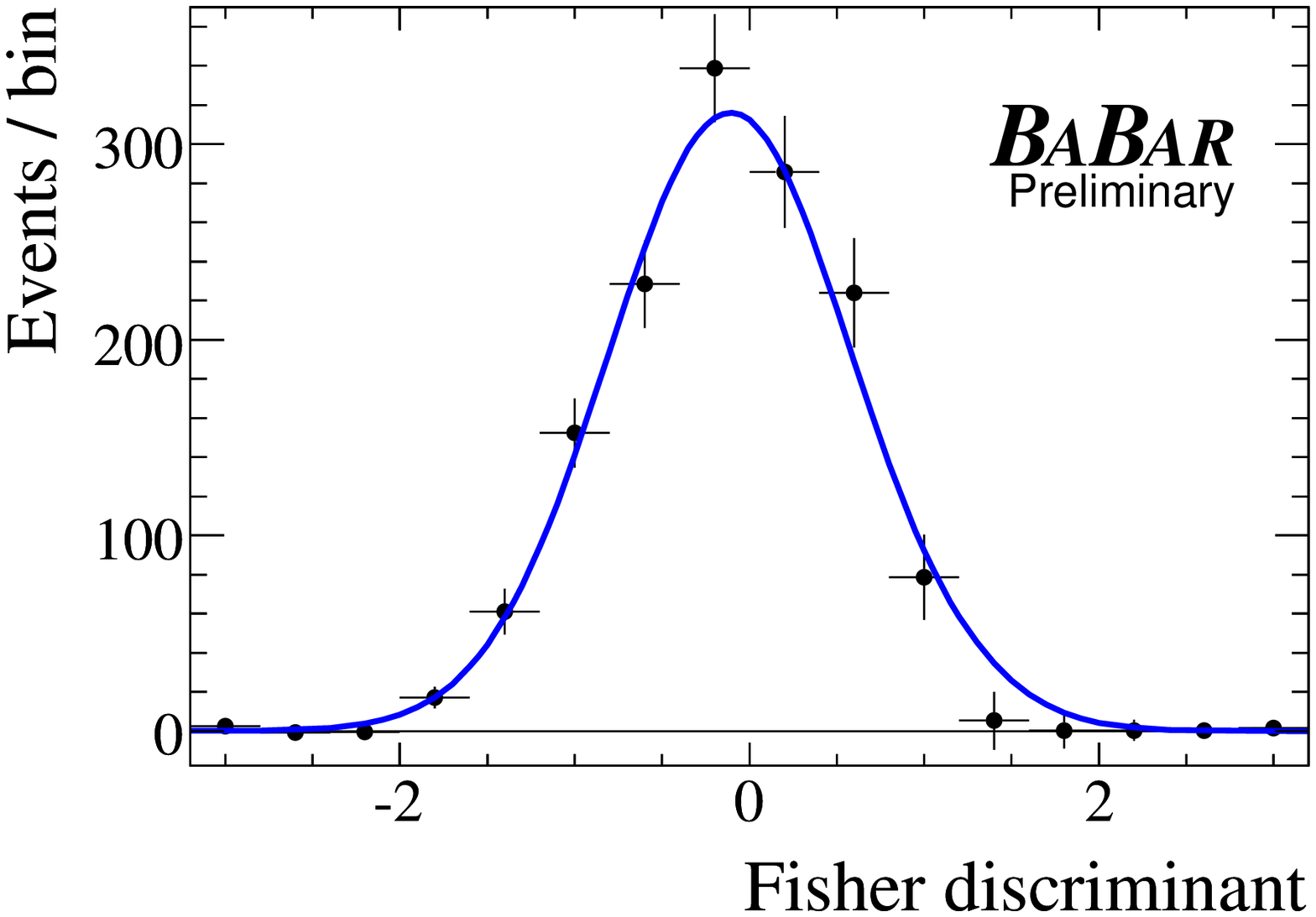} \\
\vspace*{1.5mm}
  \includegraphics[width=0.32\linewidth]{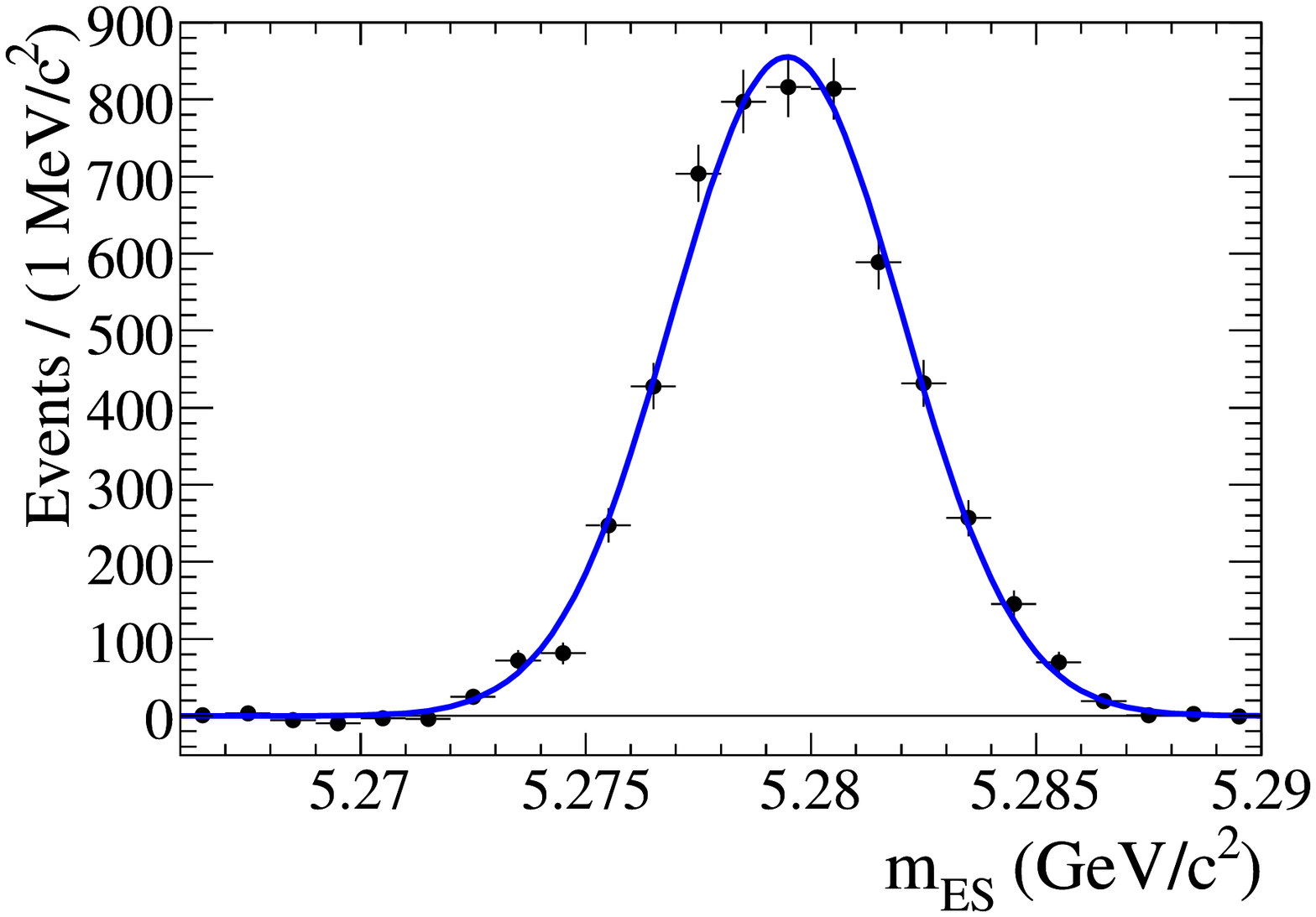}~~
  \includegraphics[width=0.31\linewidth]{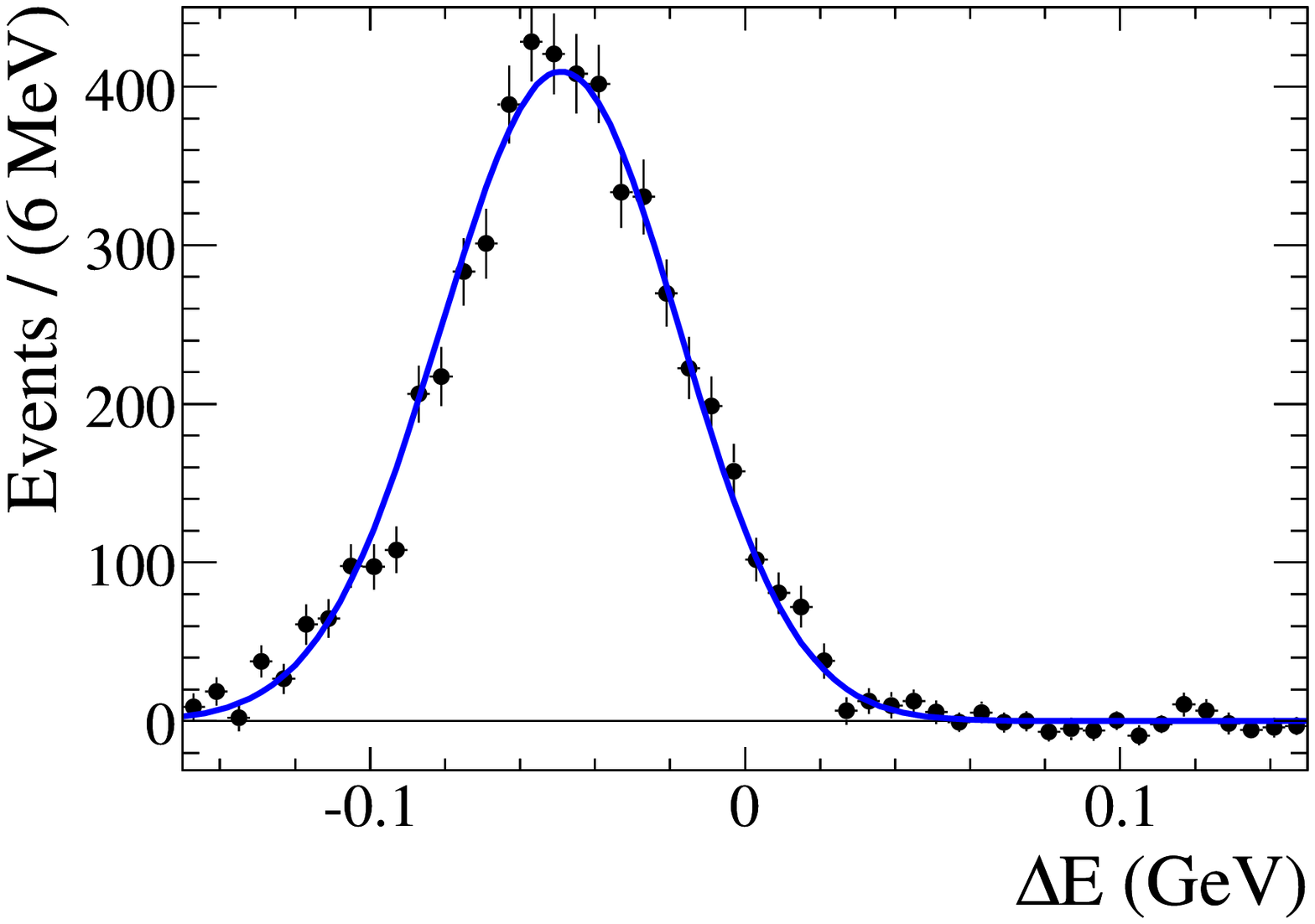}~~
  \includegraphics[width=0.31\linewidth]{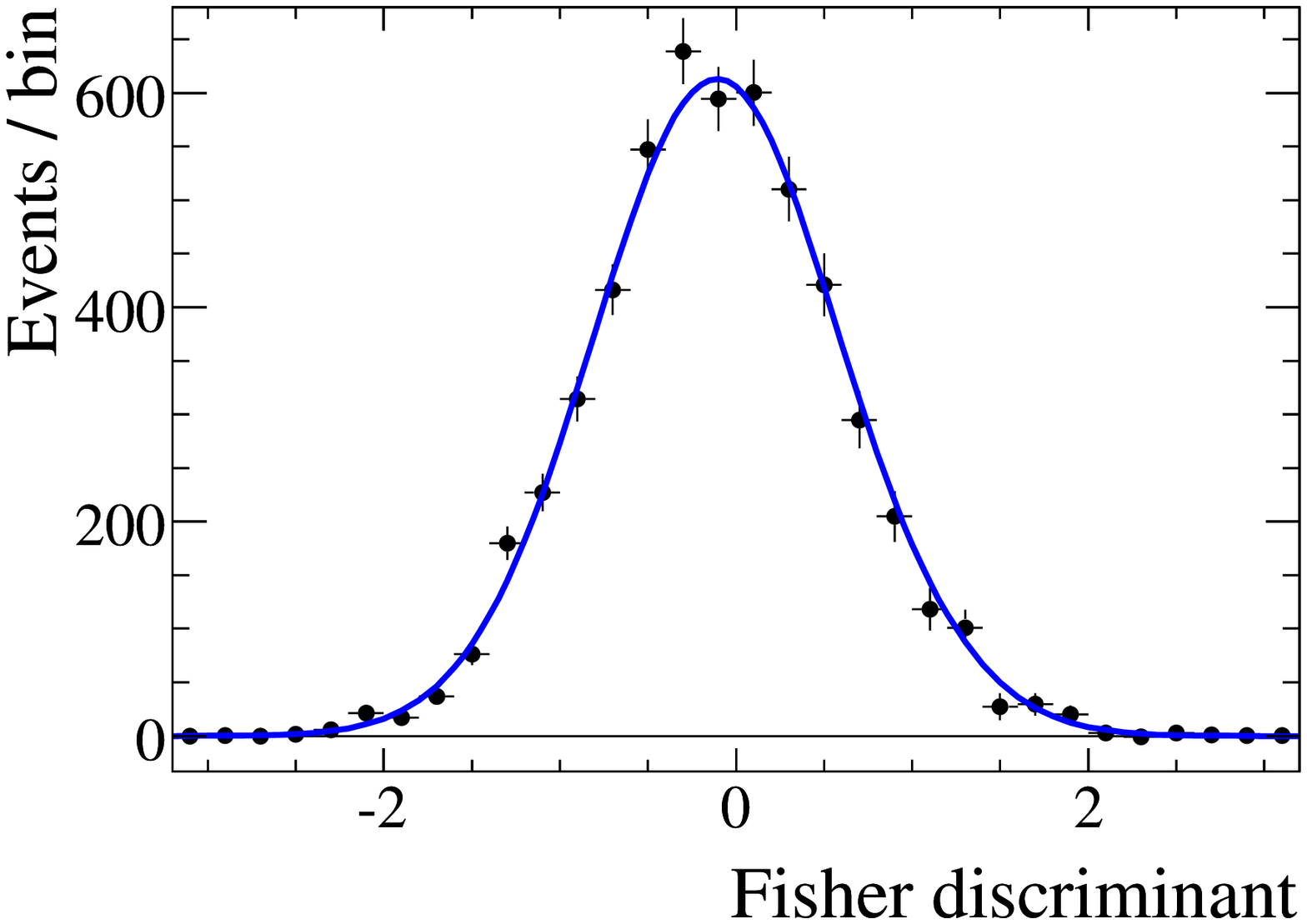} \\
\vspace*{1.5mm}
  \includegraphics[width=0.32\linewidth]{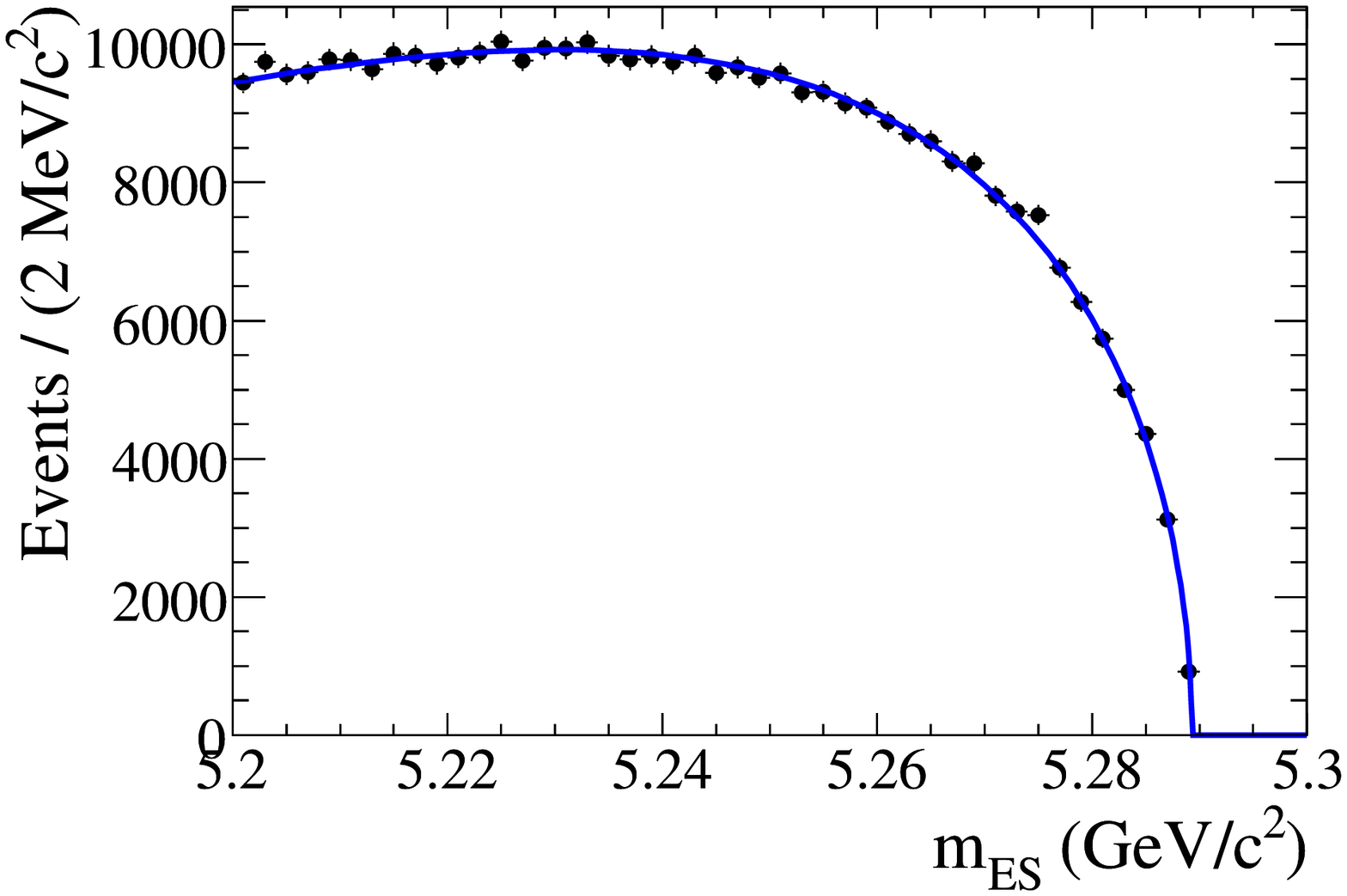}~~
  \includegraphics[width=0.31\linewidth]{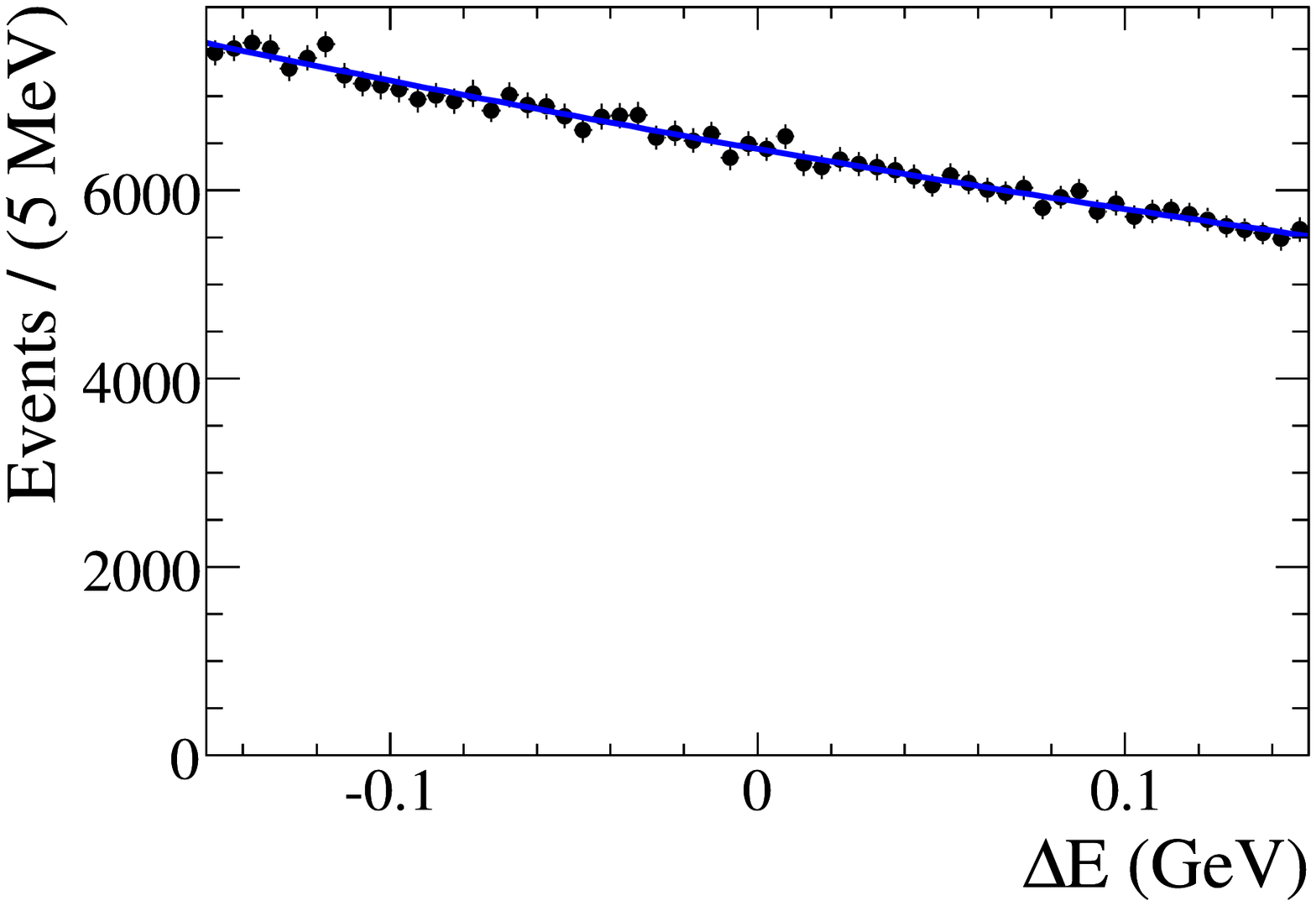}~~
  \includegraphics[width=0.31\linewidth]{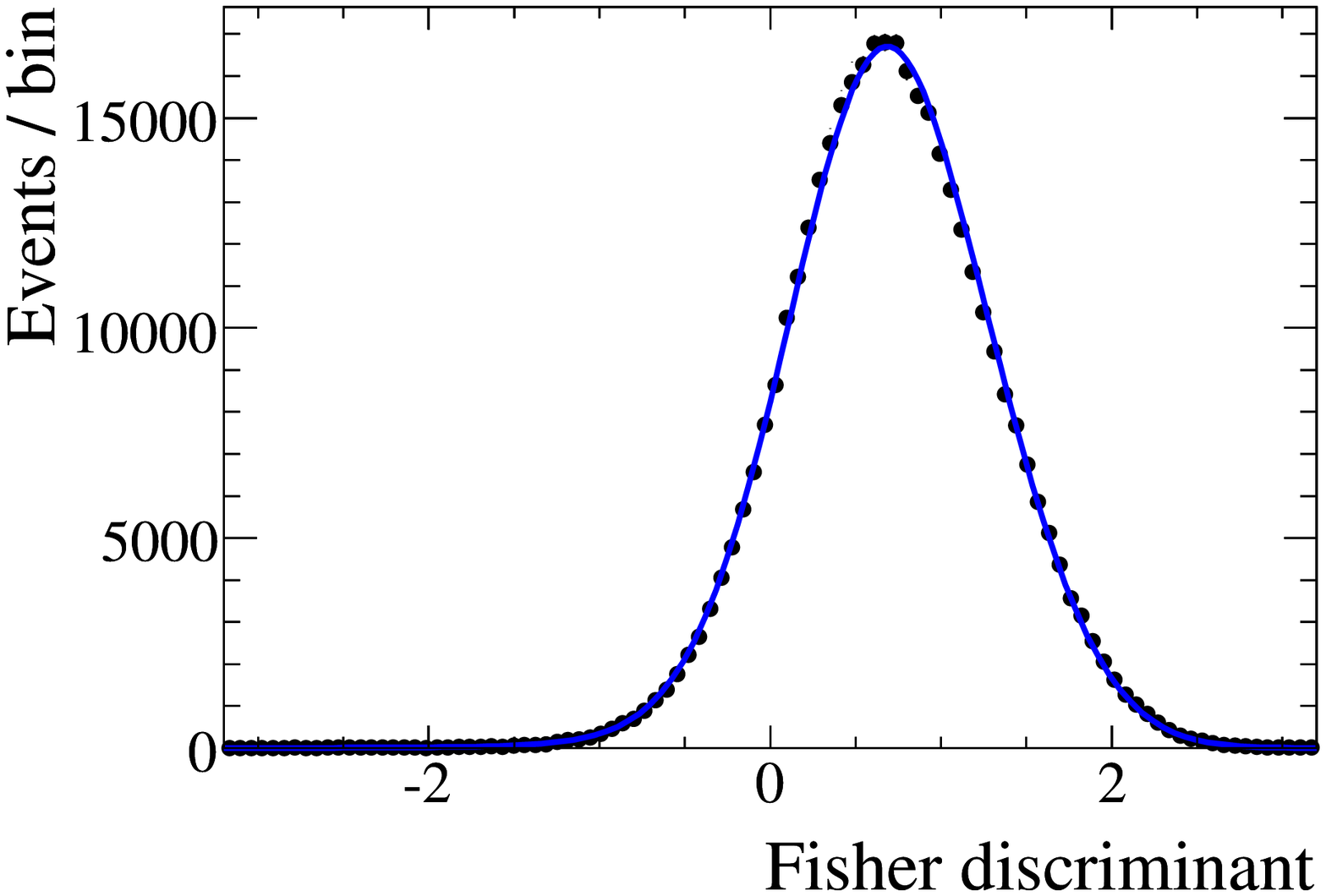}
\caption{\small The distributions of \emph{(left)} \mes, \emph{(middle)} \de, and \emph{(right)} 
Fisher discriminant \fish: 
\emph{(top)} background-subtracted for $\Bz\to\pip\pim$ signal,  
\emph{(middle)} background-subtracted for $\Bz\to\Kp\pim$ signal,  
\emph{(bottom)} signal-subtracted for all $h^+h^{\prime-}$ background candidates in the data.
The curves represent the PDFs used in the fit and reflect the fit result.
The structure to the left of the signal \DeltaE peak for $\Bz\to\pip\pim$
is consistent with the expected background from other charmless modes, which
is negligible above \unit[$-0.10$]{\gev}.
}
\label{fig:hhVar}
\end{center}
\end{figure}

\begin{figure}[!tbp]
\begin{center}
  \includegraphics[width=0.44\linewidth]{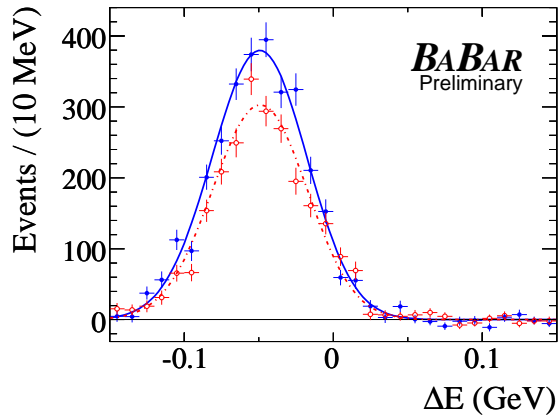}
\caption{\small  The  background-subtracted distribution of 
$\de$ for signal $\Kpm\pimp$ events, comparing \emph{(solid)} \Bz 
and \emph{(dashed)} \Bzb decays.}
\label{fig:deakpi}
\end{center}
\end{figure}

\begin{figure}[!tbph]
\begin{center}
  \includegraphics[width=0.44\linewidth]{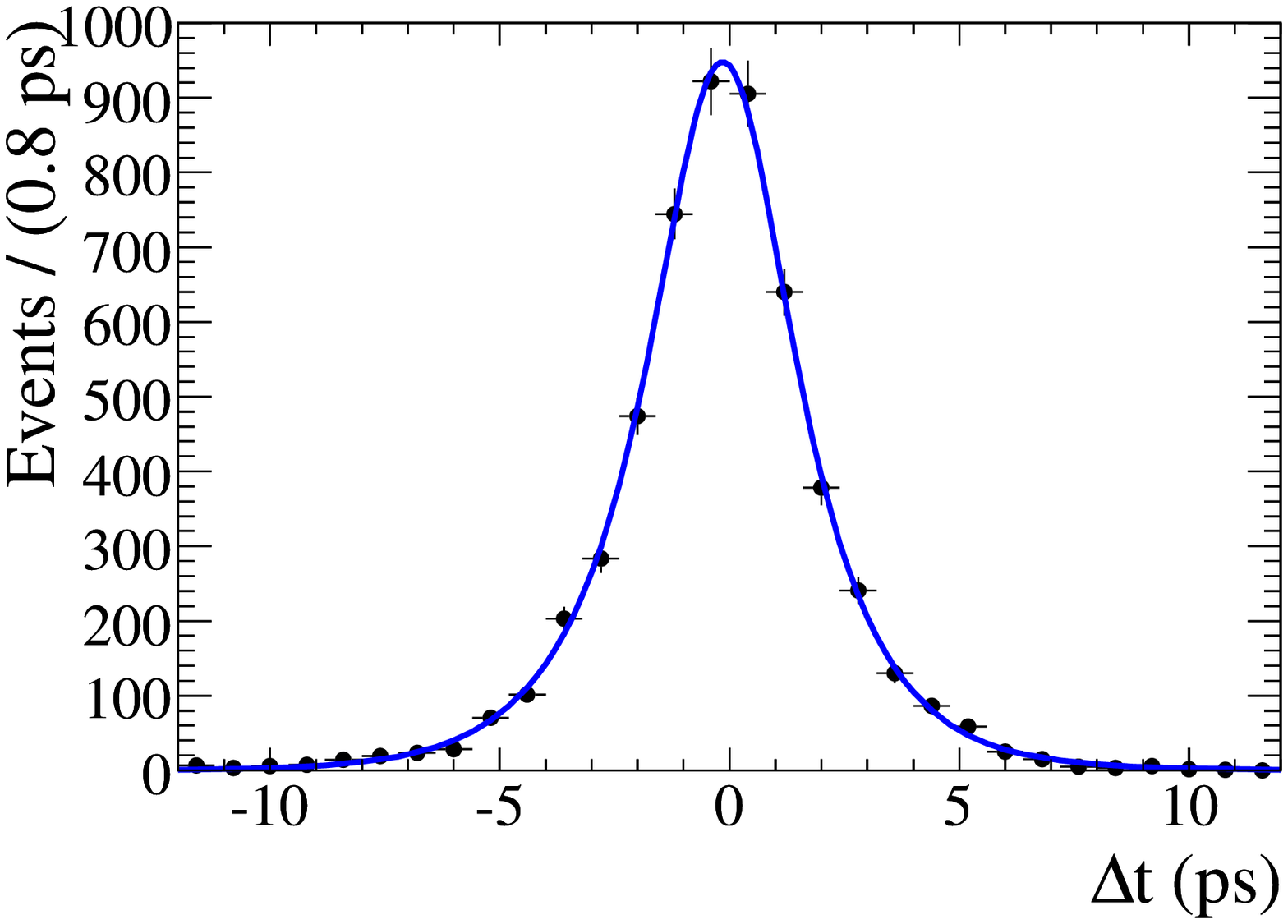}~~~
  \includegraphics[width=0.44\linewidth]{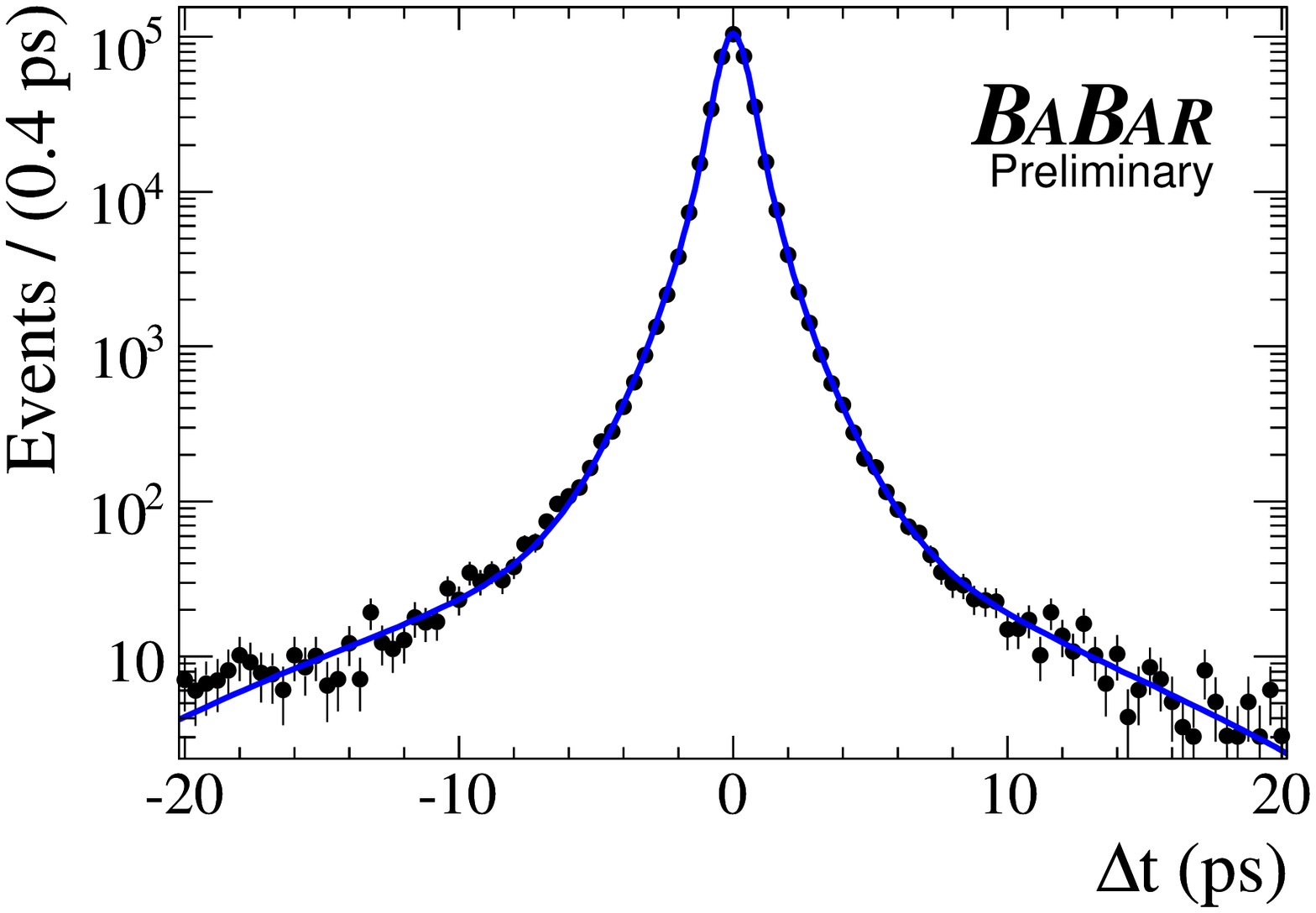}
\caption{\small   \emph{(Left)} the  background-subtracted distribution of 
$\deltat$ for  signal $\Kpm\pimp$ and 
\emph{(right)} the signal-subtracted $\deltat$ distribution for background 
candidates in the data. The curves represent the PDFs used in the fit and 
reflect the fit result.}
\label{fig:hhdt}
\end{center}
\end{figure}

\begin{figure}[!tbph]
\begin{center}
  \includegraphics[width=0.50\linewidth]{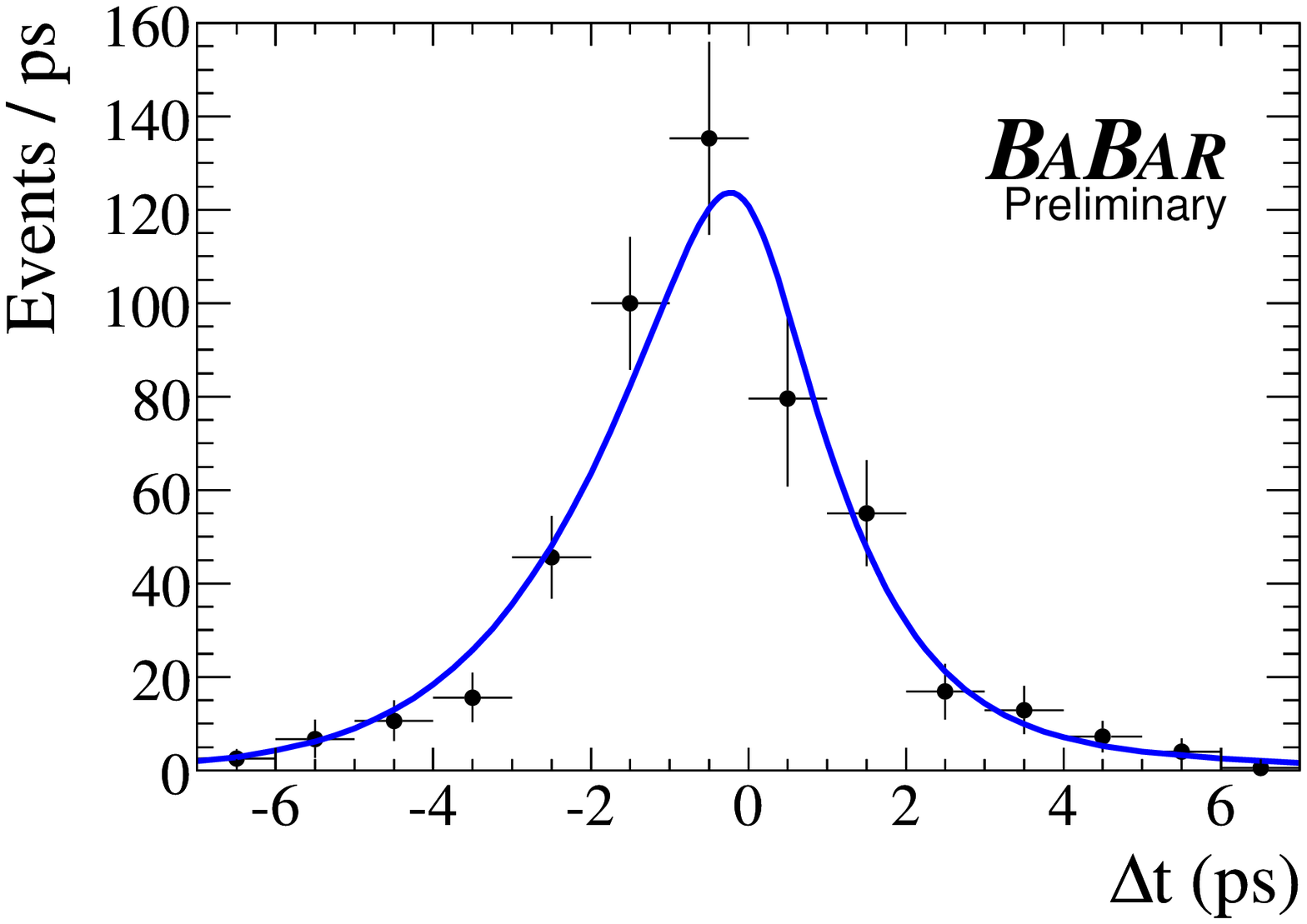} \\
\vspace*{1.5mm}
  \includegraphics[width=0.50\linewidth]{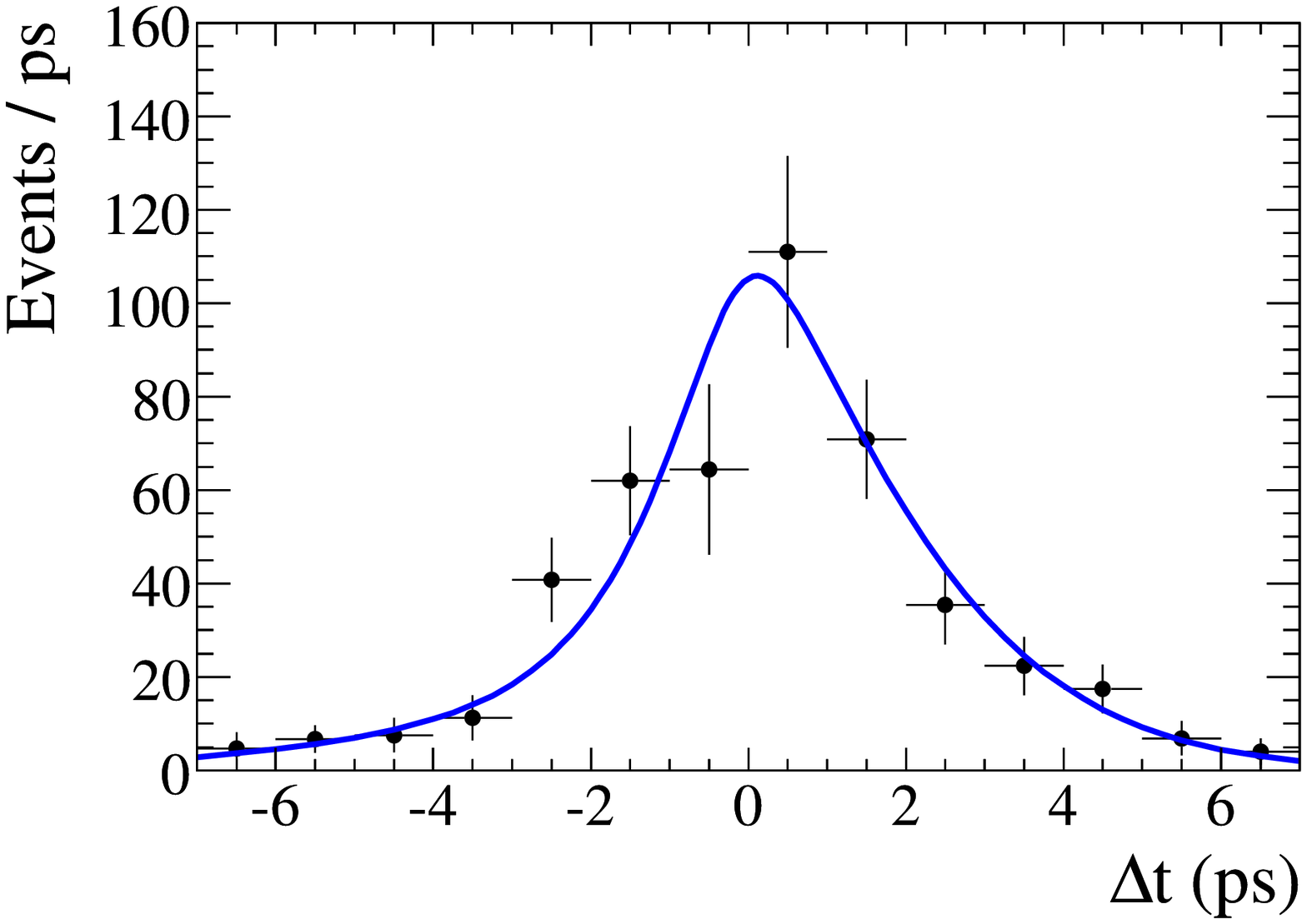} \\
\vspace*{1.5mm}
  \includegraphics[width=0.50\linewidth]{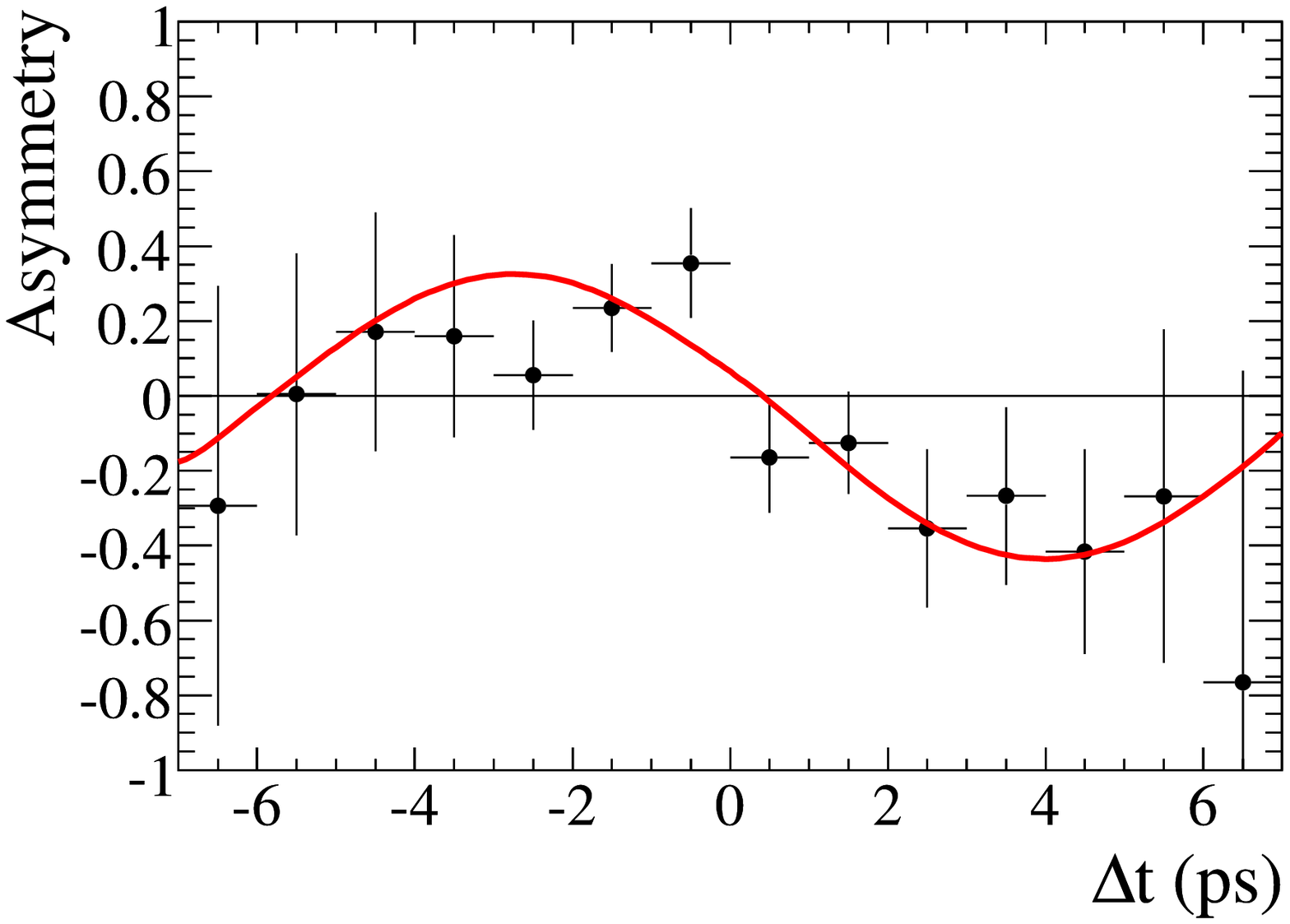}
\caption{ 
The background-subtracted distributions of 
$\deltat$ for signal $\pip\pim$ events tagged as \emph{(top)} $\Bz$  
or \emph{(middle)} $\Bzb$, and \emph{(bottom)} their asymmetry
$a(\Delta t)$ (Eq.~\ref{eq:asymmetry}).
The curves represent the PDFs used in the fit and 
reflect the fit result.}
\label{fig:asym}
\end{center}
\end{figure}

\begin{figure}[!tbp]
\begin{center}
\includegraphics[width=0.8\linewidth]{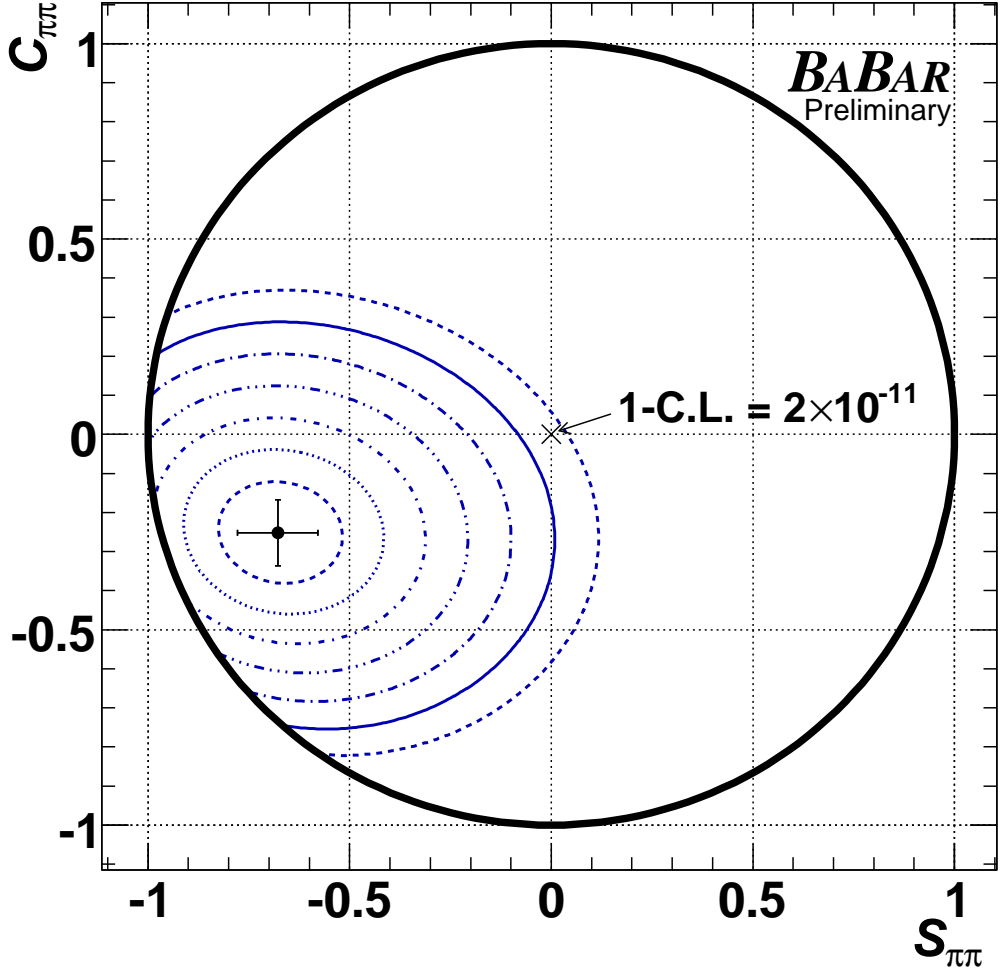}
\end{center}
\vspace{0.2cm}
\caption{ 
\spipi\ and \cpipi in $\Bz\to\pip\pim$: the central values, errors, and confidence-level (C.L.) contours for 
$1-\mathrm{C.L.} = 0.317$ $(1\sigma)$, $4.55 \times 10^{-2}$ $(2\sigma)$, $2.70 \times 10^{-3}$ $(3\sigma)$, 
$6.33 \times 10^{-5}$ $(4\sigma)$, $5.73 \times 10^{-7}$ $(5\sigma)$, $1.97 \times 10^{-9}$ $(6\sigma)$
and $2.56 \times 10^{-12}$ $(7\sigma)$,
calculated from the square root of the change in the value of
$-2\ln{\cal L}$ compared with its value at the minimum.
The systematic errors are included. The measured value is $6.7\sigma$ from the point of 
no \CP violation ($\spipi=0$, $\cpipi=0$). }
\label{fig:SCcontour}
\end{figure}

Results for the \Bztohh decay modes are listed in
Table~\ref{tab:resultsB}. The correlation coefficient between \spipi
and \cpipi is found to be $-0.056$, and 
the correlation between \cpipi and \akpi is 0.019.  
In Fig.~\ref{fig:hhVar}, we show \sPlots for $\mes$,
$\de$, and $\fish$ for the \Bztohh signal and background. 
The direct \CP asymmetry in \Bztokpi is apparent in the distribution of \de 
plotted separately for \Bz
and \Bzb decays, shown in Fig.~\ref{fig:deakpi}.  We show the
distributions of $\deltat$ for $\Bz\to\Kpm\pimp$ signal and background decays in
Fig.~\ref{fig:hhdt}. 
In Fig.~\ref{fig:asym}, we show the distribution
of \deltat separately for \Bztopipi events tagged as $\Bz$ or $\Bzb$,
and the asymmetry $a(\deltat)$.  The central values and errors for
\spipi and \cpipi are shown in Fig.~\ref{fig:SCcontour}, 
along with confidence-level contours corresponding to 
statistical significances ranging from 1 to 7 standard deviations.
Our measurement excludes the absence of  
\CP violation in \Bztopipi ($\spipi=0,\ \cpipi=0$) at a confidence level of
$2 \times 10^{-11}$, or $6.7\sigma$ (where systematic errors are taken into account). 

\begin{table}[!tbp]
\small
\caption
{\small Summary of absolute systematic errors on $\akpi$.  The total
is calculated as the quadrature sum of each contribution. To address the \akpi\ bias due 
to hadronic interactions of charged kaons with the detector material, we shift the \akpi
value obtained in the fit by $+0.0050$.}
\begin{center}
\begin{tabular}{cc}
\hline\hline
Source                           & Uncertainty\\
\hline
Material interactions            & $+0.0053$  $-0.0025$ \\
$\theta_{\rm C}$ and \dedx\ PDFs       & $0.0020$  \\
Potential MC bias                & $0.0011$  \\
Alternative DIRC parameterization & $0.0016$  \\\hline
Total                    & $+0.0060$ $-0.0037$  \\
\hline\hline
\end{tabular}
\label{tab:syst_akpi}
\end{center}
\end{table}

\begin{table}[!tbp]
\small
\caption{\small Summary of systematic uncertainties  on $\spipi$ and $\cpipi$.}
\begin{center}
\begin{tabular}{lcc}
\hline\hline
Source                     & $\spipi$ & $\cpipi$ \\
\hline
DIRC $\thetac$                  & $ 0.0064   $ & $ 0.0050   $ \\
DCH $\dedx$                   & $ 0.0032   $ & $ 0.0037   $ \\
Signal $\deltat$             & $ 0.0199   $ & $ 0.0055   $ \\
SVT local alignment        & $ 0.0004   $ & $ 0.0002   $ \\
Boost/$z$ scale            & $ 0.0021   $ & $ 0.0013   $ \\
PEP-II beam spot           & $ 0.0028   $ & $ 0.0014   $ \\
\B flavor tagging          & $ 0.0146   $ & $ 0.0138   $ \\
$\deltamd$, $\tau_{\Bz}$~\cite{pdg}   & $ 0.0004   $ & $ 0.0017   $ \\
Potential bias             & $ 0.0041   $ & $ 0.0043   $ \\
Doubly Cabibbo-suppressed decays~\cite{Owen}  & $ 0.007    $ & $ 0.016    $ \\
\hline
Total                      & $ 0.027    $ & $ 0.023   $ \\
\hline\hline
\end{tabular}
\label{tab:syst_scpipi}
\end{center}
\end{table}

Systematic uncertainties for the direct \CP asymmetry $\akpi$
are listed in Table~\ref{tab:syst_akpi}. 
Here, \akpi\ is the fitted value of the $\Kmp\pipm$ event-yield asymmetry $\akpi^{\rm raw}$ 
shifted by $+0.005^{+0.005}_{-0.003}$ to account for a bias that arises 
from the difference between the cross sections of \Kp and \Km hadronic interactions
within the \babar\ detector.  We determine this bias from a detailed MC
simulation based on GEANT4~\cite{GEANT4} version 7.1; it is independently verified with
a calculation based on the known material composition of the 
\babar\ detector~\cite{babar} and the cross sections and material properties 
tabulated in Ref.~\cite{pdg}. The corrected $\Kmp\pipm$ event-yield asymmetry in the 
background, where no observable \CP\ violation is expected, is consistent with zero:
$-0.005 \pm 0.004\, (\rm stat) ^{+0.005}_{-0.003}\, (\rm syst)$.
Systematic uncertainties for the \CP asymmetries \spipi and \cpipi
are listed in Table~\ref{tab:syst_scpipi}. They are dominated by 
uncertainties in the parameterization of
$B$-flavor tagging and vertexing, and (for $\cpipi$) in the effect of
\CP\ violation in $B_{\rm tag}$~\cite{Owen}.



\subsection{\boldmath \Bztokzpiz Results}
\label{sec:kspizResults}
\begin{figure}[!tbp]
\begin{center}
\includegraphics[width=0.8\linewidth]{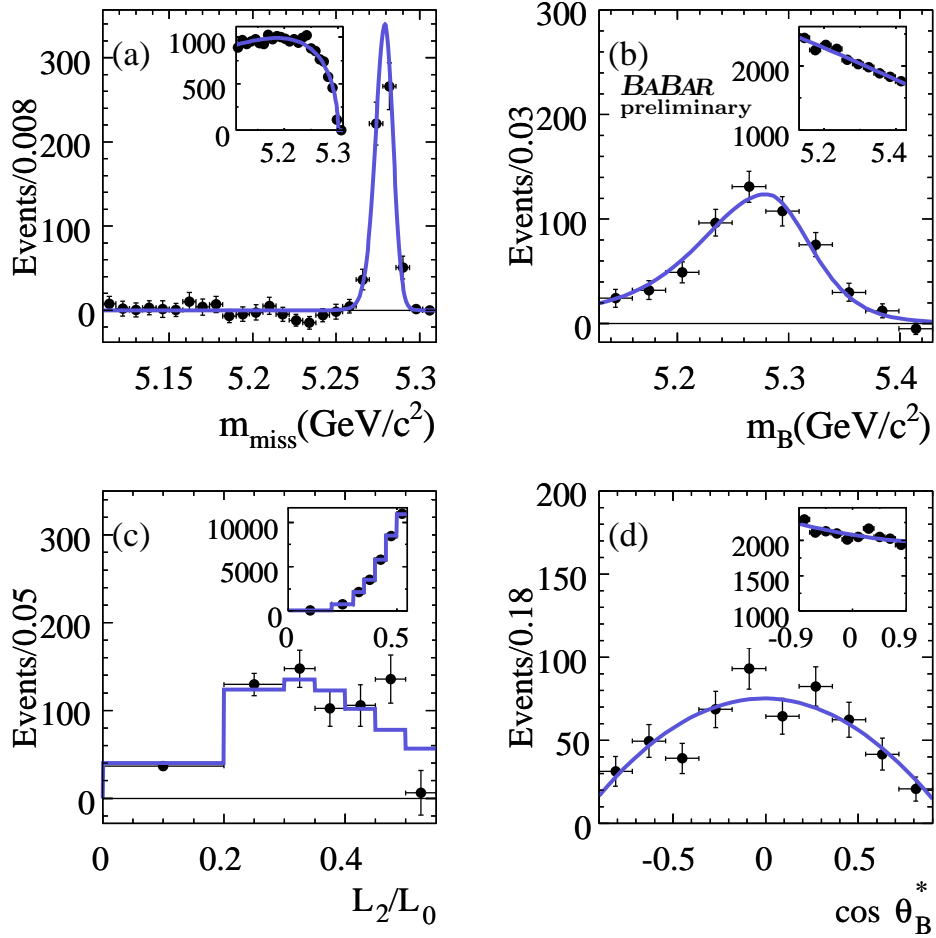}
\put(-131,323){\bfseries \babar}
\put(-131,316){\scriptsize \bfseries preliminary}
\caption{\small Distributions of \emph{(a)} $\mmiss$, \emph{(b)} $\mb$, \emph{(c)} $L_2/L_0$,
\emph{(d)} $\costhetacms$ for background-subtracted events
in the \Bztokspiz\ sample.  
The solid curves represent the shapes of signal PDFs as
obtained from the M.L.\ fit. The insets show the distributions and 
PDFs for signal-subtracted data. 
}
\label{fig:splots}
\end{center}
\end{figure}

Results for the \Bztokzpiz decay mode are
summarized in Table~\ref{tab:resultsA}. 
In Fig.~\ref{fig:splots}, we show \sPlots for $m_{\rm miss}$, $m_B$, 
$L_2/L_0$, and $|\costhetacms|$ for signal events, with background distributions 
shown in the insets. 

To compute the systematic error associated with the statistical
precision on the parameters of the likelihood function, we shift
each parameter by its associated uncertainty and repeat the fit.
For \deltat and the tagging parameters, the uncertainty is obtained
from the fit to the \Bflav\ sample, while for the other parameters it
is obtained from MC; the total error is obtained by summing the
individual contributions in quadrature.  This fit systematic also
accounts for the limited statistics available to determine the shape
of the likelihood function in Eq.~\ref{eq:ml}.  We find a systematic error of
$1.2$ events on the $\KS\piz$ yield.  As an additional systematic error
associated with the data--MC agreement of the shape of the
signal PDFs, we also quote the largest deviation observed when the
parameters of the individual signal PDFs for \mmiss, \mb, $L_2/L_0$,
and \costhetacms{} are floated in the fit. This gives a systematic
error on the yield of $2.5$ events.  The output values of the PDF
parameters are also used to assign a systematic error to the selection
efficiency of the cuts on the likelihood variables.  Comparing the
efficiency in data to that in the MC, we obtain a relative systematic error of
$1.5\%$. We do not assign a systematic uncertainty on the scale of
\mmiss{} and \mb{} because we float these variables in the fit.  We evaluate the
systematic error due to the neglected correlations among fit
variables using a set of MC experiments, in which we embed signal
events from a full detector simulation with events generated from the
background PDFs. Since the shifts are small and only marginally
significant, we use the average shift in the yield ($+2.2$ events) as
the associated systematic uncertainty.

We estimate the background from other $B$ decays to be small in the
nominal fit. We account for a systematic shift induced on the signal
yield by this neglected component by embedding simulated $B$ background
events in the data set and evaluating the average shift in the fit
result: $+5.2$ events on the signal yield. We adjust the signal yield
accordingly and use half of the shift as a systematic uncertainty.

For the branching fraction, additional systematic errors come from the 
uncertainty in the 
selection efficiency, 
the counting of \BB{} pairs in the data sample
(1.1\%),
and the branching fractions in the \Bz decay chain, ${\cal B}(K^0_S
\to \pi^+\pi^-)=0.6920 \pm 0.0005$ and ${\cal B}(\pi^0 \to \gamma
\gamma) = 0.98798 \pm 0.00032$~\cite{pdg}. The systematic uncertainties 
are summarized in Table~\ref{tab:BFsys}.

\begin{table}[!tbp]
\small
    \caption{\small Summary of dominant contributions to the systematic error
    on the measurement of ${\cal B}(\Bztokzpiz)$ \label{tab:BFsys}}
  \begin{center}
    \begin{tabular}{c|cc}
      \hline\hline           &                                                 & $\sigma_{\rm syst}({\cal B})/{\cal B}$ (\%) \\
      \hline  
\multirow{3}{*}{Efficiencies}&      $\pi^0$ efficiency                         & $3.0$ \\
                             &      $\KS$ efficiency                           & $0.5$ \\
                             &      Cut on likelihood variables                & $1.5$ \\
      \cline{1-1}
\multirow{5}{*}{Yield}       &      stat.\ precision on PDF parameters          & $0.22$ \\
                             &      Shape of signal PDFs                        & $0.45$ \\
                             &      $\BB$ background                  & $0.47$  \\
                             &      Correlations among likelihood variables     & $0.40$  \\
                             &      Resolution function                        & $0.49$  \\
      \cline{1-1}
Normalization                &      Number of $\BB$ pairs                            & $1.1$  \\
      \hline
      Total                  &                                                 & $3.7$  \\
      \hline\hline
    \end{tabular}
  \end{center}
\end{table}

\section{CONCLUSIONS}
\label{sec:Conclusions}

\begin{figure}[tbp]
\begin{center}
\includegraphics[width=0.69\linewidth]{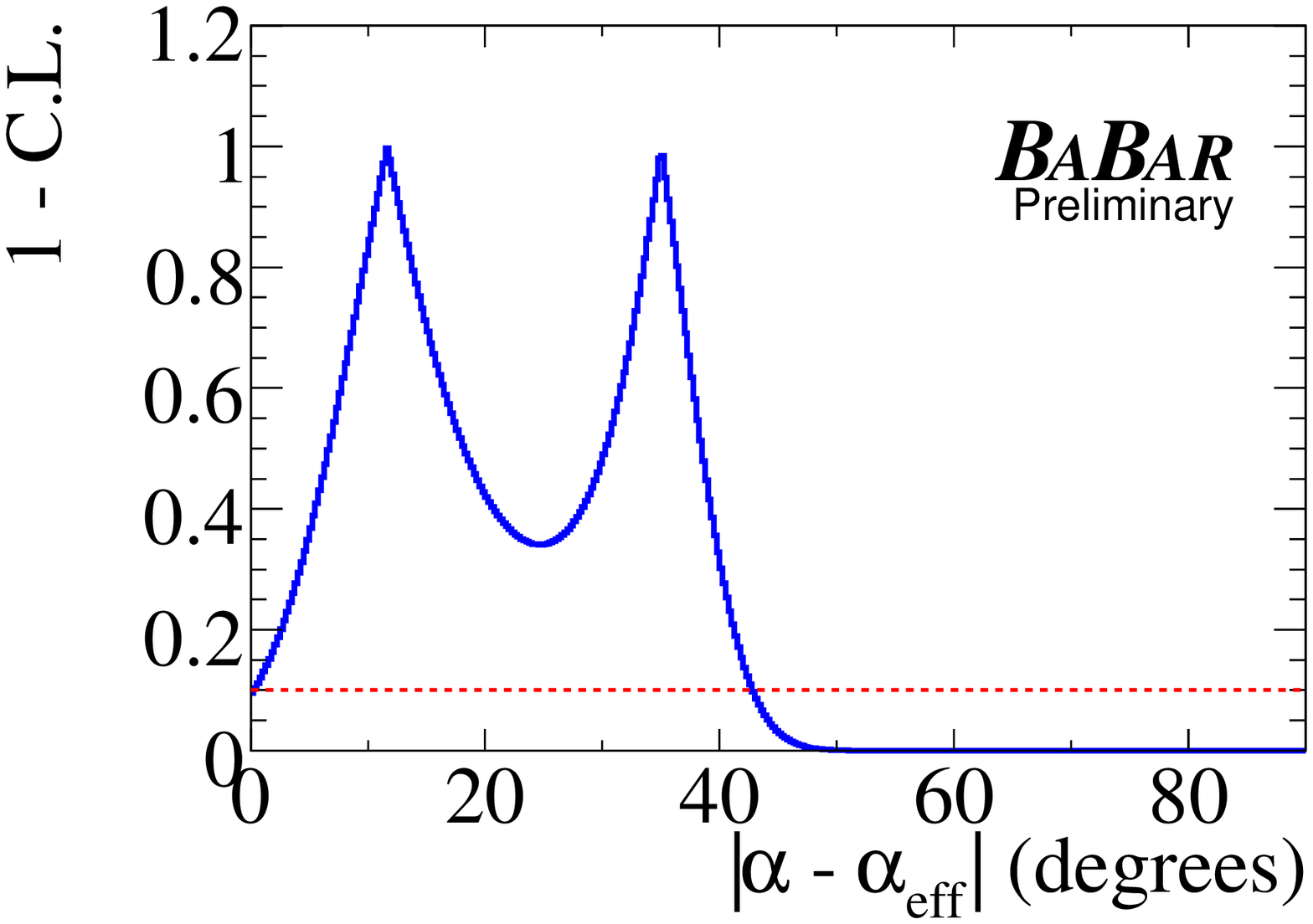}
\includegraphics[width=0.69\linewidth]{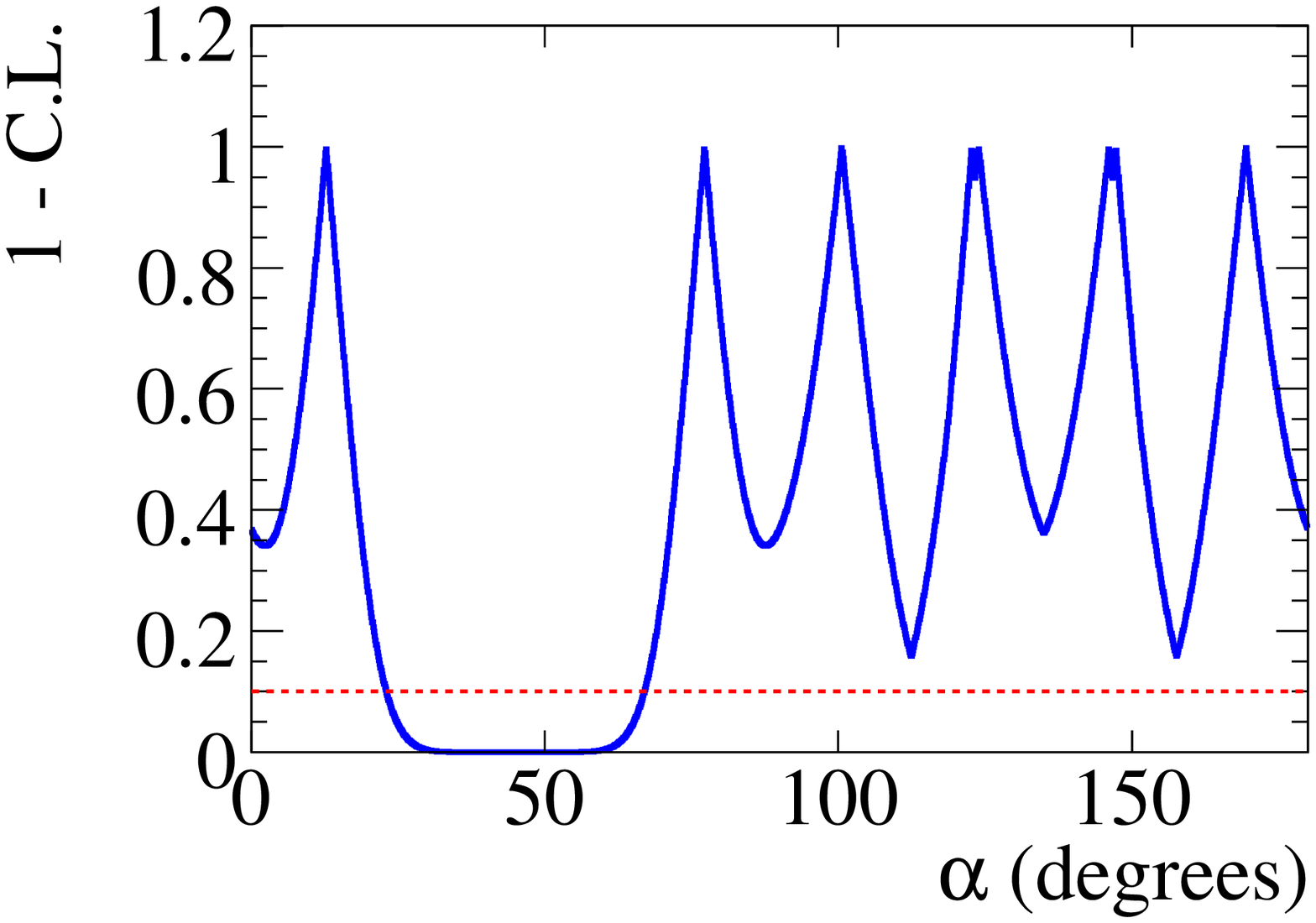}
\end{center}
\vspace{0.1cm}
\caption{
 \emph{(Top)}
  constraint on the angle $\delalph = \alpha-\alphaeff$,
  expressed as one minus the confidence level (C.L.), as a function of
  $|\delalph|$.  We find an upper bound on $\left|\delalph\right|$ of
  $43^{\rm o}$ at the 90\% C.L. 
\emph{(Bottom)}
 constraint on the CKM angle $\alpha$
  expressed as $1-$C.L. There are eight peaks, two of them
  nearly merged, corresponding to an eight-fold ambiguity in the
  extraction of $\alpha$; four solutions are from the value and sign of
  $\delalph$, which is doubled due to the trigonometric reflections
  between $\alphaeff$ and $\pi/2 - \alphaeff$. 
  We exclude the range $[23^{\rm o}, 67^{\rm o}]$ in $\alpha$ at the 90\% C.L.
Only the isospin-triangle relations and the expressions in
  Eq.~\ref{eq:asymmetry} are used in this constraint. 
  The point $\alpha=0$, which corresponds to no \CP violation, and
  the values of $\alpha$ near 0 or $\pi$ can be excluded with additional
  physics input~\cite{pi0pi0_BaBar,UTFit_2007_pipi}.
}
\label{fig:alpha}
\end{figure}


The \CP-asymmetry and branching-fraction results described in this
paper are:
\begin{align*}
   \spipi & =   -0.68 \pm 0.10 \pm 0.03, \\
   \cpipi & =   -0.25 \pm 0.08 \pm 0.02, \\
   {\cal A}_{K\pi} & = -0.107 \pm 0.016 ^{+0.006}_{-0.004}, \\
   \cpizpiz & =  -0.43 \pm 0.26 \pm 0.05, \\
   \BR(\Bztopizpiz) & = ( 1.83 \pm 0.21 \pm 0.13 ) \times 10^{-6}, \\
   \BR(\Bztokzpiz) & = ( 10.1 \pm 0.6 \pm 0.4 ) \times 10^{-6}. 
\end{align*}
We combine \BR(\Bztopizpiz) with the branching fractions
$\BR(\Bztopippim) = (5.5 \pm 0.4 \pm 0.3) \times 10^{-6}$ 
and $\BR(B^{\pm}\to\pi^{\pm}\pi^0) = (5.02 \pm 0.46 \pm 0.29)\times 10^{-6}$ 
previously measured by \babar~\cite{pi0pi0_BaBar, BabarBRPiPi} to evaluate the
constraints on both the penguin contribution to $\alpha$ and on the
CKM angle $\alpha$ itself. Constraints are evaluated by scanning the
parameters of interest, $\left|\delalph\right| = |\alpha - \alphaeff|$ and $\alpha$, and
then calculating the $\chi^{2}$ for the five amplitudes ($A^{+0}$,
$A^{+-}$, $A^{00}$, $\Abar^{+-}$, $\Abar^{00}$) from our
measurements and the isospin-triangle relations~\cite{ref:CKMfitter}. The $\chi^2$ is
converted to a confidence level (C.L.) as shown in Fig.~\ref{fig:alpha}.  
The upper bound on  $\left|\delalph\right|$ is $43^{\rm o}$ at the 90\% C.L.,
and the range $[23^{\rm o}, 67^{\rm o}]$ in $\alpha$ 
is excluded at the 90\% C.L.
If we consider only the solution preferred in the SM~\cite{Gronau2007},
$\alpha$ is in the range $[71^{\rm o},109^{\rm o}]$ at the 68\% C.L. 
Somewhat more restrictive new constraints on $\alpha$ have been found in the
measurements of $B\to\rho\rho$ and $\Bz\to(\rho\pi)^0$ decays~\cite{ref:rhorho}.

We have also presented an improved measurement of
the \CP-violating charge asymmetry \akpi\ in the $\Bz\to\Kp\pim$ decay.
We observe direct \CP violation in $\Bz\to\Kp\pim$ with a 
significance of $6.1\sigma$.
Ignoring color-suppressed tree amplitudes, the charge asymmetries in $\Kp\pim$ 
and $\Kp\piz$ should be equal (see Gronau and Rosner in Ref.~\cite{ref:SumRule}), which has 
not been supported by recent \babar\ and Belle data~\cite{BaBarPRL2007,pi0pi0_BaBar,BelleNature2008}.  
These results might
indicate a large color-suppressed amplitude, an enhanced electroweak penguin,
or possibly new-physics effects~\cite{ref:NP}.

Finally, we have presented an improved measurement of
\BR(\Bztokzpiz).  From the rate sum-rule prediction~\cite{ref:SumRule}
$2 \BR(K^0\piz)^{\rm sr} =  \BR(K^+\pim) + \frac{\tau_0}{\tau_+}[\BR(K^0\pip)
- 2 \BR(K^+\piz) ]  $ and the currently published results for the
other three $B\ra K\pi$ modes, we find the sum-rule prediction to be
$\BR(\Bztokzpiz)^{\rm sr} = (8.4 \pm 0.8)\times 10^{-6}$, which is
consistent with our new experimental result.

\section{ACKNOWLEDGMENTS}
\label{sec:Acknowledgments}

We are grateful for the 
extraordinary contributions of our \pep2\ colleagues in
achieving the excellent luminosity and machine conditions
that have made this work possible.
The success of this project also relies critically on the 
expertise and dedication of the computing organizations that 
support \babar.
The collaborating institutions wish to thank 
SLAC for its support and the kind hospitality extended to them. 
This work is supported by the
U.S.\ Department of Energy
and National Science Foundation, the
Natural Sciences and Engineering Research Council (Canada),
the Commissariat \`a l'Energie Atomique and
Institut National de Physique Nucl\'eaire et de Physique des Particules
(France), the
Bundesministerium f\"ur Bildung und Forschung and
Deutsche Forschungsgemeinschaft
(Germany), the
Istituto Nazionale di Fisica Nucleare (Italy),
the Foundation for Fundamental Research on Matter (The Netherlands),
the Research Council of Norway, the
Ministry of Education and Science of the Russian Federation, 
Ministerio de Educaci\'on y Ciencia (Spain), and the
Science and Technology Facilities Council (United Kingdom).
Individuals have received support from 
the Marie-Curie IEF program (European Union) and
the A. P. Sloan Foundation.




\begin{thebibliography}{99}
\bibitem{refConj}
  The use of charge-conjugate modes is implied throughout this paper unless otherwise noted.
\end{thebibliography}
\end{document}